\documentclass{article}
\usepackage{subfiles}

\usepackage{amsmath}
\usepackage{amssymb}
\usepackage{siunitx}
\DeclareSIUnit\Molar{\textsc{m}}
\usepackage{dblfloatfix}
\usepackage{fixltx2e}

\usepackage{gensymb}
\usepackage[affil-it]{authblk}
\usepackage{color}
\usepackage{rotating}
\usepackage{xcolor}
\DeclareSIUnit\Molar{\textsc{m}}

\usepackage{graphicx}

\usepackage{cite}
\begin{document}

\newcommand*\samethanks[1][\value{footnote}]{\footnotemark[#1]}

\title{Modeling the ballistic-to-diffusive transition in nematode motility reveals variation in exploratory behavior across species}

\author[1]{Stephen J. Helms\thanks{These authors contributed equally to this work.}}
\affil[1]{AMOLF Institute, Amsterdam, The Netherlands}

\author[1]{W. Mathijs Rozemuller\samethanks}
 
\author[2]{Antonio Carlos Costa\samethanks}
\affil[2]{Dept. of Physics \& Astronomy, Vrije Universiteit, Amsterdam, The Netherlands}

\author[3]{Leon Avery}
\affil[3]{Dept. of Physiology and Biophysics, Virginia Commonwealth Univ., Richmond, VA, USA}

\author[2,4]{Greg J. Stephens}
\affil[4]{Okinawa Institute of Science and Technology, Onna-son, Okinawa, Japan}

\author[1]{Thomas S. Shimizu
\thanks{Electronic address: \texttt{shimizu@amolf.nl}; Corresponding author}}

\date{}
\maketitle
%7896 words
\begin{abstract}
{A quantitative understanding of organism-level behavior requires predictive models that can capture the richness of behavioral phenotypes, yet are simple enough to connect with underlying mechanistic processes. Here we investigate the motile behavior of nematodes at the level of their translational motion on surfaces driven by undulatory propulsion. We broadly sample the nematode behavioral repertoire by measuring motile trajectories of the canonical lab strain \textit{C. elegans} N2 as well as wild strains and distant species. We focus on trajectory dynamics over timescales spanning the transition from ballistic (straight) to diffusive (random) movement and find that salient features of the motility statistics are captured by a random walk model with independent dynamics in the speed, bearing and reversal events. We show that the model parameters vary among species in a correlated, low-dimensional manner suggestive of a common mode of behavioral control and a trade-off between exploration and exploitation.  The distribution of phenotypes along this primary mode of variation reveals that not only the mean but also the variance varies considerably across strains, suggesting that these nematode lineages employ contrasting ``bet-hedging'' strategies for foraging.}

\end{abstract}

\section*{Introduction}
A ubiquitous feature of biological motility is the combination of stereotyped movements in seemingly random sequences. Capturing the essential characteristics of motion thus requires a statistical description, in close analogy to the random-walk formulation of Brownian motion in physics. A canonical example is the ``run-and-tumble'' behavior of \textit{E. coli} bacteria, in which relatively straight paths (runs) are interspersed by rapid and random reorientation events (tumbles) \cite{Berg1972}. The random walk of \textit{E. coli} can thus be characterized by two random variables (run length and tumble angle) and two constant parameters (swimming speed and rotational diffusion coefficient), and detailed studies over decades have yielded mechanistic models that link these key behavioral parameters to the underlying anatomy and physiology \cite{Lovely1975, Schnitzer1990, DeGennes2004, Celani2011}. Random-walk theory has been fruitfully applied also to studies of eukaryotic cell migration in both two \cite{Gail1970,Tranquillo1988,Selmeczi2005} and three \cite{Wu2014} dimensions.

Can a similar top-down approach be fruitfully applied to more complex organisms--for example, an animal controlled by a neural network? Animal behavior is both astonishing in its diversity and daunting in its complexity, given the inherently high-dimensional space of possible anatomical, physiological, and environmental configurations. It is therefore essential to identify appropriate models and parameterizations to succinctly represent the complex space of behaviors --- a non-trivial task that has traditionally relied on the insights of expert biologists. In this study, we ask if one can achieve a similar synthesis by an alternative, physically-motivated approach \cite{Brown2018}. We seek a quantitative model with predictive power over behavioral statistics, and yet a parameterization that is simple enough to permit meaningful interpretations of phenotypes in a reduced space of variables. As an example, we focus on the motile behavior of nematodes, which explore space using a combination of random and directed motility driven by undulatory propulsion.

The nematode \textit{C. elegans} has long been a model organism for the genetics of neural systems \cite{Brenner1974,bargmann2013}, and recent advances in imaging have made it feasible to record a large fraction of the worm's nervous system activity at single-cell resolution \cite{Kato2015,Venkatachalam2016,Nguyen2016}. These developments raise the compelling possibility of elucidating the neural basis of behavioral control at the organism scale, but such endeavors will require unambiguous definitions of neural circuit outputs and functional performance. The worm's behavioral repertoire \cite{Gjorgjieva2014,Cohen2014a} is commonly characterized in terms of forward motion occasionally interrupted by brief reversals \cite{Croll1975,Croll1975a,Roberts2016}, during which the undulatory body wave that drives its movement \cite{Gray1964} switches direction. In addition, worms reorient with a combination of gradual curves in the trajectory (``weathervaning'') \cite{Iino2009a,Peliti2013} and sharp changes in body orientation (omega-turns \cite{Croll1975a} and delta-turns \cite{Broekmans2016}). These elementary behaviors are combined in exploring an environment \cite{Pierce-Shimomura1999,Iino2009a}. Environmental cues such as chemical, mechanical, or thermal stimuli \cite{Faumont2012} lead to a biasing of these behaviors, guiding the worm in favorable directions \cite{Pierce-Shimomura1999,Iino2009a,Ryu2002}. Finally, in practical terms, the worm's small size ($\sim$\SI{1}{\milli\meter} in length), moderate propulsive speed ($\sim$\SI{100}{\micro\meter\per\second}) and short generation time ($\sim$2 days) allow a considerable fraction of its behavioral repertoire to be efficiently sampled in the laboratory \cite{Croll1975,Fujiwara2002}. An influential example of such an analysis is the ``pirouette'' model proposed by Pierce-Shimomura and Lockery \cite{Pierce-Shimomura1999} which describes the worm's exploratory behavior as long runs interrupted occasionally by bursts of reversals and omega turns that reorient the worm, in close analogy to the run-and-tumble model of bacterial random walks \cite{Berg1972}.  Later work by Iino et al. identified that worms also navigate by smoother modulations of their direction during long runs (``weathervaning'') \cite{Iino2009a}, and Calhoun et al. have suggested that \textit{C. elegans} may track the information content of environmental statistics in searching for food \cite{Calhoun2014}, a motile strategy that has been termed 'infotaxis' \cite{Vergassola2007}. A recent study by Roberts et al. \cite{Roberts2016} analyzed high (submicron) resolution kinematics of \textit{C. elegans } locomotion and developed a stochastic model of forward-reverse switching  dynamics that include the  short-lived ($\sim$\SI{0.1}{\second}) pause states that were identified between forward and reverse runs.

Importantly, while these previous studies have illuminated different modes of behavioral control, they were not designed to obtain a predictive model of the trajectory statistics and thus a succinct parameterization of \textit{C. elegans} motility remains an important open problem. A quantitative parameterization capturing the repertoire of \textit{C. elegans}' behavioral phenotypes would facilitate data-driven investigations of behavioral strategies: for example, whether worms demonstrate distinct modes of motility (characterized by correlated changes in parameters) over time, or in response to changes in environmental conditions \cite{Fujiwara2002,Gallagher2013,Flavell2013,Salvador2014}. Variation in the obtained parameters among individuals can inform on the distribution of behavioral phenotypes within a population, and reveal evolutionary constraints and trade-offs between strategies represented by distinct parameter sets \cite{Shoval2012}. 

 \textit{C. elegans} is a member of the \textit{Nematoda} phylum, one of the largest and most diverse phylogenetic groups of species \cite{DeLey2006,Corsi2015}. Despite the diversity of ecological niches these animals inhabit \cite{DeLey2006}, comparisons of nematode body plans have revealed a remarkable degree of conservation, even down to the level of individual neurons \cite{Rabinowitch2008}.  This combination of highly conserved anatomy and ecological diversity makes nematode motility a compelling case for studies of behavioral phenotypes. Anatomical conservation suggests it might be possible to describe the behavior of diverse nematodes by a common model, and identifying the manner in which existing natural variation is distributed across the parameter space of the model could reveal distinct motility strategies resulting from optimization under different environmental conditions.

In this study, we develop a simple random walk model describing the translational movements of a diverse collection of nematode species, freely-moving on a two-dimensional agar surface.  In addition to providing a quantitative and predictive measure of trajectory dynamics, the parameters of our model define a space of possible behaviors. Variation within such a space can occur due to changes in individual behavior over time (reflecting temporal variation in the underlying sensorimotor physiology, or ``mood''), differences in behavior among individuals (reflecting stable differences in physiology, or ``personality'') and differences between strains and species (reflecting cumulative effects of natural selection). By quantitative analyses of such patterns of variation, we seek to identify simple, organizing principles underlying behavior.

\section*{Results}

\subsection*{Nematodes Perform Random Walks Off-Food with a Broad Range of Diffusivities Across Strains}

\begin{figure*}
\centerline{\includegraphics[width=16cm]{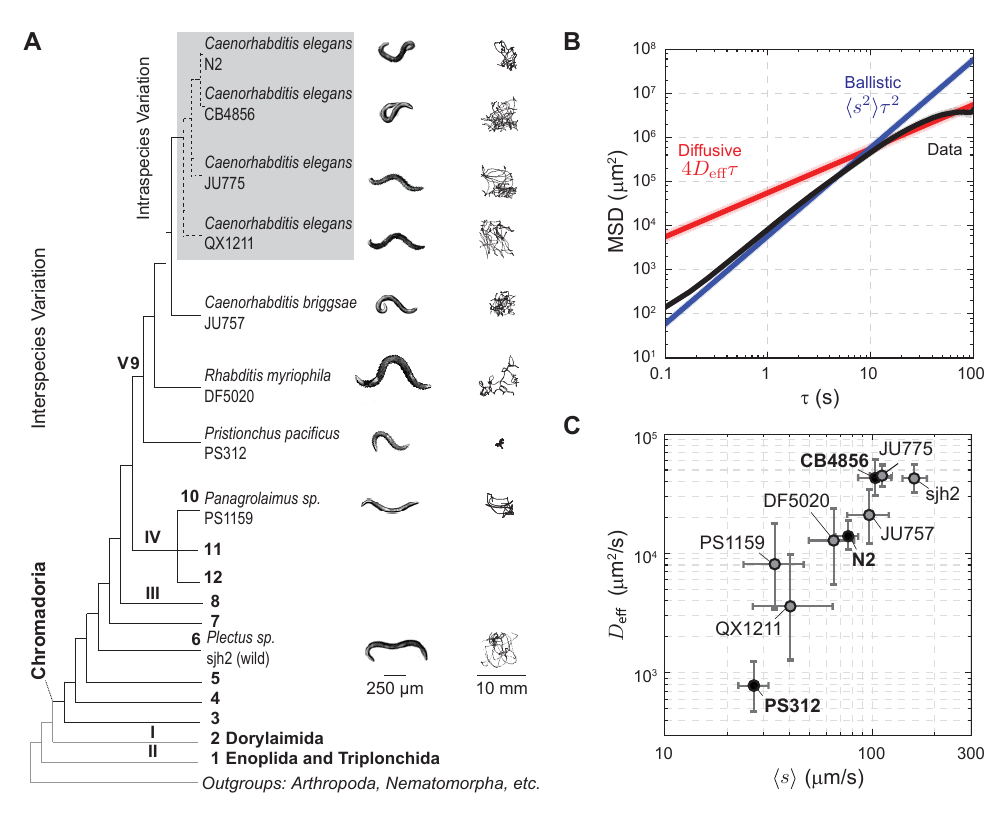}}
\caption{
{Nematodes perform random walks off-food with a mean speed and effective diffusivity that varies across strains.}  
(A) Phylogenetic tree with the strains used in this study.
The bold numbers are the major clades of \textit{Nematoda}.
The gray box indicates genetically distinct wild isolates of \textit{C. elegans}.
A representative worm image and 30 minute trajectory are shown to the right. Shaded regions indicate a 95\% confidence interval.
(B) The average mean-squared displacement, MSD, across N2 individuals is shown in black.
For comparison, we show the MSD expected from ballistic (blue) and diffusive (red) dynamics.
The motility transitions from a ballistic to diffusive regime within a time scale of tens of seconds.
(C) Mean speed $\langle s\rangle$ and effective diffusivity $D_\text{eff}$ (mean and 95\% confidence intervals) for each strain, calculated from fits of the mean-squared displacement as in B. Across strains, both $\langle s\rangle$ and $D_\text{eff}$ vary by orders of magnitude.
}
\label{Figure 1}
\end{figure*}

In order to identify conserved and divergent aspects of motility strategies, we sampled motile behavior over a broad evolutionary range. We selected a phylogenetically diverse collection of nematodes with an increased sampling density closer to the laboratory strain \textit{C. elegans} (Figure \ref{Figure 1}A and Supplementary Information, SI). To sample individual variation, we recorded the motility of up to 20 well-fed individuals per strain and each individual for 30 minutes on a food-free agar plate at \SI{11.5}{\hertz} with a resolution of \SI{12.5}{\micro\metre}/px (see SI).

We measured the centroid position ($\vec{x}(t)$) and calculated the centroid velocity ($\vec{v}(t)$), using image analysis techniques (Figure S11 and SI). We chose the centroid as the measure of the worm's position because it effectively filters out most of the dynamics of the propulsive body wave. There was considerable variation in the spatial extent and degree of turning visible in the trajectories both within and across strains (Figure \ref{Figure 1}A, S2). 

As previously seen in \textit{C. elegans} \cite{Stephens2010}, the measured mean-squared displacement,
\begin{equation}\label{MeanSquareDisplacement}
\langle [\Delta x(\tau)]^2 \rangle \equiv \langle |\vec{x}(t+\tau)-\vec{x}(t)|^2\rangle,
\end{equation}

\noindent
revealed a transition from ballistic to diffusive motion within a \SI{100}{\second} timescale (Figure \ref{Figure 1}B, S3). Over short times, the worm's path was relatively straight, with the mean-squared displacement scaling quadratically with the time lag $\tau$ and speed $s$ as $\langle s^2 \rangle \tau^2$ (\textit{i.e.} a log-log slope of 2). Over longer times, the slope decreased with $\tau$ reflecting the randomization of orientation characteristic of diffusion, and an effective diffusivity $D_\text{eff}$ was extracted by fits to $\langle\left[\Delta x(\tau)\right]^ 2\rangle = 4D_{\text{eff}}\tau$ (see SI). For times $\gtrsim \SI{100}{\second}$, the slope of the mean-squared displacement decreased yet further due to encounters of the worm with walls of the observation arena. We confirmed that this confinement did not affect detection of the ballistic to diffusive transition  (Figure S1). This analysis revealed that the visible differences in the spatial extent of these 30-minute trajectories stem from variation by nearly an order of magnitude in speed and two orders of magnitude in diffusivity (Figure \ref{Figure 1}C, Tables S1 \& S2). 

\subsection*{The Random Walk of Nematodes Can Be Decomposed into Speed, Turning and Reversal Dynamics}

The broad range of observed speeds and diffusivities suggest that these diverse nematodes have evolved a variety of strategies for spatial exploration. To gain further insights into the manner in which such contrasting behaviors are implemented by each strain, we sought to extract a minimal model of the nematodes' random walk by further decomposing the trajectory statistics of all nine measured strains. In this and the following three sections, we illustrate our analysis and model development with data from three contrasting strains: CB4856 and PS312, which demonstrated two of the most extreme phenotypes, and the canonical laboratory strain N2 (see SI for equivalent data for all strains).

The translational motion of the worm can be described by the time-varying centroid velocity $\vec{v}(t)$ which can in turn be decomposed into speed $s(t)$ and direction of motion (hereafter referred to as its ``bearing'') $\phi(t)$:

\begin{equation}\label{VelocityComponents}
\vec{v}(t) = \dfrac{d\vec{x}(t)}{dt} = s(t) \left[ \cos \phi(t), \sin \phi(t)\right]
\end{equation}

\begin{figure}
\centerline{\includegraphics[width=11.4cm]{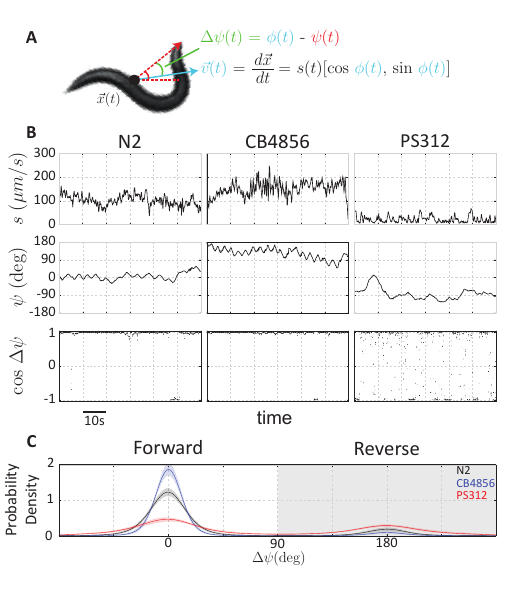}}
\caption{
{The random walk of nematodes is composed of speed, turning, and reversal dynamics.}
(A) We describe the motility of the worm by the time-varying quantities $s(t)$ (speed; black), $\psi(t)$ (orientation; red), and $\Delta\psi(t)$ (alignment; green) which measures the difference between the alignment of the velocity $\phi(t)$ (blue) and $\psi(t)$.
(B) One minute examples of speed, orientation, and velocity alignment time series for individuals from three exemplar strains.
(C) The probability distribution of $\Delta\psi(t)$ reveals bimodality corresponding to forward and reverse motion. Shaded regions indicate a 95\% confidence interval.
}
\label{Figure 2}
\end{figure}

To account for head-tail asymmetry in the worm's anatomy, we additionally define the body orientation ($\psi(t)$; hereafter referred to simply as ``orientation'') by the angle of the vector connecting the worm's centroid to the head (Figure \ref{Figure 2}A). The centroid bearing is related to this orientation of the worm by
\begin{equation}\label{Bearing}
\phi(t) = \psi(t) + \Delta\psi(t)
\end{equation}
where the difference $\Delta{\psi}(t)$ is a measure of the alignment of the direction of movement with the worm's body orientation (hereafter referred to simply as ``alignment''). We found for all strains that the distribution of $\Delta{\psi}(t)$ was bimodal with peaks at 0\degree and 180\degree (Figure \ref{Figure 2}C, S7A). These match the forward and reverse states of motion described in \textit{C. elegans} \cite{Croll1975,Croll1975a}.

Each of the three components of the worm's motility (speed, orientation, and alignment) varied considerably over time and in qualitatively different ways between strains (Figure \ref{Figure 2}B). For example, the three strains shown in Figure \ref{Figure 2}B differed not only in their average speed, but also in the amplitude and timescale of fluctuations about the average speed. Similarly, the statistics of orientation fluctuations about the drifting mean also differed visibly between strains. Finally, transitions between forward and reverse runs were far more frequent in PS312 as compared to N2 and CB4856. Given the apparently random manner in which these motility components varied over time, we proceeded to analyze the dynamics of each of these three components as a stochastic process.

\subsection*{Speed Dynamics}
Speed control has not been extensively studied in \textit{C. elegans}, but it is known that worms move with a characteristic speed that is influenced by stimuli \cite{Faumont2012}. When intervals corresponding to transitions between forward and reverse runs were excluded from the time series, we found that the autocorrelation in speed fluctuations decayed exponentially over a few seconds (Figure \ref{Figure 3}A, S5A), a timescale similar to the period of the propulsive body wave.
These dynamics are naturally captured by an Ornstein-Uhlenbeck process \cite{Kampen2007}, which describes random fluctuations arising from white noise (increments of a diffusive Wiener process, $dW_t$ \cite{Kampen2007}) with magnitude $\sqrt{2 D_s}$ that relax with timescale $\tau_s$ back to an average value, $\mu_s=\langle s \rangle$:
\begin{equation}\label{SpeedDynamics}
ds(t) = \tau_s^{-1} \left[ \mu_s - s(t) \right] dt + \sqrt{2 D_s} dW_t
\end{equation}
Numerical integration of this equation closely reproduced the observed speed distributions during runs (Figure S5B).

\subsection*{Diffusive Turning with Drift}
\begin{figure*}
\centerline{\includegraphics[height=15cm]{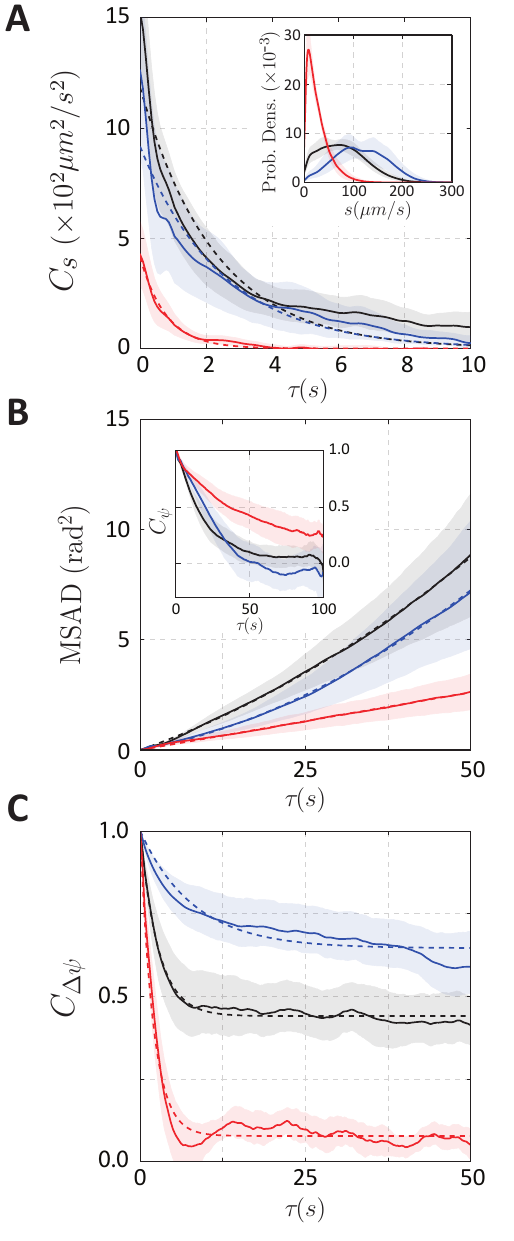}}
\caption{
{Statistical characterization of the motility dynamics.}
(A) The autocorrelation of the speed indicated that fluctuations decayed exponentially over a few seconds.
(A, inset) Speed distributions for three exemplar strains.
(B) The mean-squared angular displacement (MSAD) increased quadratically.
(B, inset) The orientation autocorrelation function did not decay exponentially, with some worms demonstrating significant undershoots below zero.
(C) The velocity alignment autocorrelation decayed exponentially over tens of seconds to a positive constant. 
In each plot, the ensemble average for all individuals from the strains are shown with solid lines and trends are shown with dashed liens. Shaded regions indicate a 95\% confidence interval.
}
\label{Figure 3}
\end{figure*}

The orientation $\psi(t)$ captures turning dynamics that are independent of abrupt changes in bearing $\phi(t)$ due to reversals.  To change orientation, \textit{C. elegans} executes a combination of large, ventrally-biased \cite{Gray2005a} sharp turns \cite{Croll1975,Broekmans2016} and gradual ``weathervaning'' \cite{Iino2009a}, both of which contribute to randomization of orientation over time. This random walk in orientation was not purely diffusive: the orientation correlation $C_\psi(\tau) =  \langle \cos\left[\psi(t+\tau)-\psi(t)\right]\rangle $ does not decay exponentially (Figures \ref{Figure 3}B Inset, S6B), and the mean-squared angular displacement, $\text{MSAD}(\tau)=\langle [ \psi(t+\tau)-\psi(t) ]^2 \rangle$, increases nonlinearly with time (Figures \ref{Figure 3}B, S6A).

We found that this nonlinear MSAD of $\psi(t)$ could be well fit by a quadratic function of the time delay $\tau$:  $\text{MSAD}(\tau) = k_{\psi \mbox{rms}}^2\tau^2+2 D_{\psi} \tau$, corresponding to a diffusion-and-drift model with root-mean-square (rms) drift magnitude $k_{\psi \mbox{rms}}$ and angular diffusion coefficient $D_\psi$ (see Supporting Information for derivation). A non-zero drift magnitude $k_{\psi \mbox{rms}} \neq 0$  indicates that in addition to purely random (diffusive) changes in orientation, there is an underlying bias (i.e. directional persistence) in the worms' turning over \SI{100}{\s} windows, consistent with previous studies in larger arenas \cite{Peliti2013}.

These observations lead to a simple model for the orientation dynamics that combines drift (approximated as a deterministic linear process over a \SI{100}{\s} window) with stochastic diffusion: 
\begin{equation}\label{OrientationDynamics}
d\psi(t) = k_\psi dt + \sqrt{2D_\psi} dW_t,
\end{equation}
where we set the drift magnitude $k_\psi=k_{\psi\mbox{rms}}$ and $dW_t$ represents increments of a Wiener process \cite{Kampen2007}.

We note that while this model described well the orientation dynamics within \SI{100}{\second} windows, over longer timescales additional dynamics may be relevant.  The magnitude of $k_\psi$ in our data ($\sim$\SI{1}{\degree\per\second}) was similar to that of weathervaning excursions reported for  \textit{C. elegans} navigating in salt gradients \cite{Iino2009a}.

\subsection*{Forward and Reverse Runs}

The observation that motion during runs switched abruptly between forward and reverse states (with $\Delta\psi\approx\{0\degree,180\degree\}$, respectively; Figures \ref{Figure 2}B,C,S7A) suggested that reversals could be described as a discrete stochastic process.
The manner in which reversals contribute to randomization of bearing over a time lag $\tau$ is captured by the autocorrelation function of $\Delta\psi (t)$,  $C_{\Delta\psi}(\tau) \equiv \langle \cos(\Delta \psi(t+\tau)-\Delta \psi(t)) \rangle$. We found that $C_{\Delta\psi}(\tau)$ decayed nearly exponentially to a non-zero baseline (Figure \ref{Figure 3}C, Figure S7C). This is the predicted behavior for the autocorrelation function of the simplest of two-state processes (a ``random telegraph process''):
\begin{equation}\label{ForwardRunDistribution}
P(T_{\text{fwd}}>t) = \exp(-t/\tau_{\text{fwd}})
\end{equation}
\begin{equation}\label{ReverseRunDistribution}
P(T_{\text{rev}}>t) = \exp(-t/\tau_{\text{rev}}),
\end{equation}
\noindent in which the distribution of forward and reverse run intervals ($T_\text{fwd}$ and $T_\text{rev}$) are completely determined by a single time constant ($\tau_\text{fwd}$ and $\tau_\text{rev}$, respectively). The random telegraph process yields an autocorrelation function that decays exponentially as $C_{\Delta\psi}(\tau)=C_{\Delta\psi}(\infty)+\left(1-C_{\Delta\psi}(\infty)\right)e^{-\tau/\tau_{RT}}$ to a minimum value $ C_{\Delta\psi}(\infty) \equiv ( (\tau_\text{fwd}-\tau_\text{rev})/(\tau_\text{rev} + \tau_\text{fwd}))^2  $ with a timescale $\tau_{RT}\equiv \left( \tau_\text{fwd}^{-1} + \tau_\text{rev}^{-1} \right)^{-1} $ \cite{Papoulis1984}. Results obtained from fitting the autocorrelation function are consistent with those obtained from the distribution of time intervals between detected switching events (figure S7, SI).
In principle, the forward and reverse states could be characterized by differences in motility parameters of our model other than these transition times, as forward and reverse motion are driven by distinct command interneurons in \textit{C. elegans} \cite{Chalfie1985,Piggott2011}. However, we found that run speeds were nearly identical between forward and reverse runs (Figure S8). While we expect that this symmetry will be broken under some specific conditions, such as the escape response \cite{Culotti1978a}, the strong speed correlation between the two states motivates the assumption, adopted in our model, that reversals change only the bearing (by 180\degree) and the propensity to reverse direction, represented in our model by the time constants $\tau_{\text{fwd}}$ and $\tau_{\text{rev}}$. 

\subsection*{A Model with Independent Speed, Turning and Reversals Captures the Ballistic-to-Diffusive Transition in Nematode Motility}

\begin{figure*}
\centerline{\includegraphics[scale=1]{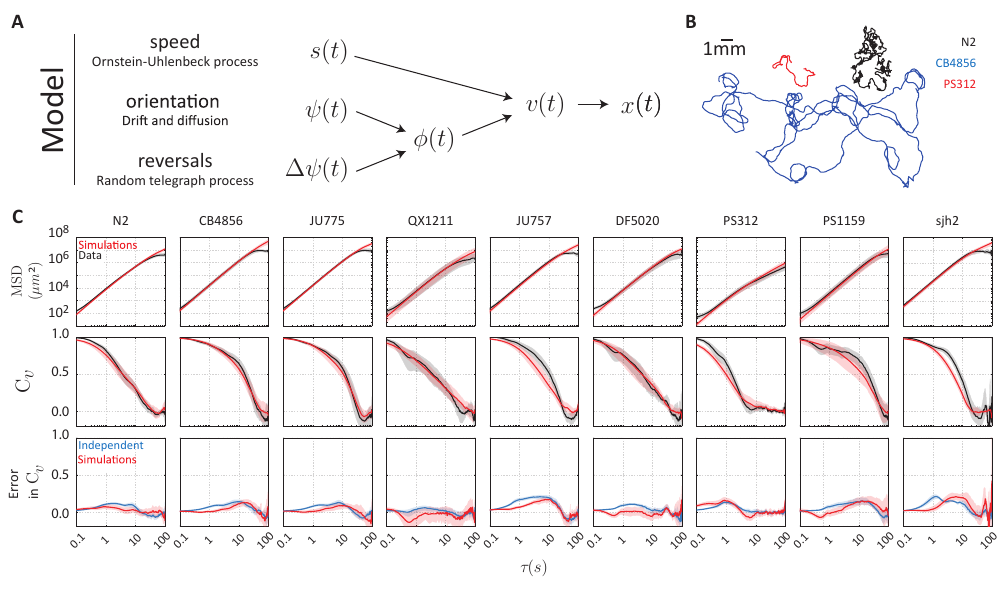}}
\caption{
A model consisting of independent speed (Ornstein-Uhlenbeck process), turning (drift and diffusion), and reversal dynamics (random telegraph process) quantitatively captures nematode motility.
(A) Summary of the model.
(B) Simulated trajectories for the three exemplar strains.
(C) Statistical comparison of the data (black) and simulations (red), ensemble averaged across individuals for each strain.
(C, top) The mean-squared displacement (MSD) was closely reproduced in all cases.
(C, middle) The normalized velocity autocorrelation, $C_{\vec{v}}(\tau)/C_{\vec{v}}(0)$, (VACF) was less well captured.
(C, bottom) The relatively small errors in the simulated VACF (red) can be traced to the assumption of independence in the dynamics of the speed, orientation, and velocity alignment (blue). Shaded regions indicate a 95\% confidence interval.
}
\label{Figure 4}
\end{figure*}

Given that the dynamics of the worm's speed, turning and reversals could be described as simple stochastic processes, we asked whether combining them as independent components in a model of the worms' random walk could sufficiently describe the observed motility statistics (Figure \ref{Figure 4}A). We simulated trajectories of worms by numerically integrating equations \eqref{SpeedDynamics}-\eqref{ReverseRunDistribution} for the speed, orientation, and reversal dynamics, respectively, which yields the worm's velocity dynamics through equations \eqref{VelocityComponents} and \eqref{Bearing}, with $\Delta\psi(t)$ equal to 0\degree\,during forward runs and 180\degree\,during reverse runs. Simulations of this model using parameters fit to individual worms produced trajectories that qualitatively resembled real trajectories and varied considerably in their spatial extent (Figure \ref{Figure 4}B).

Next, we quantitatively assessed the performance of the model in reproducing the statistics of the observed trajectories over the time scale of 100 s, within which all strains completed the transition from ballistic to diffusive motion  (Figure \ref{Figure 4}C). We found that the model based on independent speed, turning and reversal dynamics closely reproduced not only the diffusivity of each strain but also the time evolution of the mean-squared displacement ($\langle [\Delta x(\tau)]^2 \rangle$) across the ballistic-to-diffusive transition (Figure \ref{Figure 4}C, top).  A closer inspection of the dynamics across this transition is possible by examining the velocity autocorrelation function ($C_{\vec{v}}(\tau)$), the time integral of which determines the slope of the mean-squared displacement through $(d/dt)\langle [\Delta x(\tau)]^2\rangle=2\int_{0}^{\tau}{d\tau' C_{\vec{v}}(\tau')}$, a variant of the Green-Kubo relation \cite{Green1954,Kubo1957}. The transition from ballistic to diffusive motion is characterized by the manner in which the normalized velocity autocorrelation $C_{\vec{v}}(\tau)/C_{\vec{v}}(0)$ decays over the time lag $\tau$ from unity (at $\tau=0$) to zero (as $\tau \to \infty$). We found that $C_{\vec{v}}(\tau)$ varied considerably across strains, not only in the overall ballistic-to-diffusive transition time, but also in the more detailed dynamics of the autocorrelation decay over time (Figure \ref{Figure 4}C, middle). Salient features, such as the transition time, of the measured velocity autocorrelation functions $C_{\vec{v},\text{obs}}$ were reproduced closely by the simulated velocity autocorrelation functions $C_{\vec{v},\text{model}}$, but there were also subtle deviations in the detailed dynamics for a number of strains.

Given our model's simplifying assumption that dynamics for $s(t)$, $\psi(t)$, and $\Delta\psi(t)$ are independent stochastic processes, we asked whether the remaining discrepancies between the simulated and measured velocity autocorrelation dynamics could be explained by violations of this assumption of independence. As a model-free assessment of the degree of non-independence, we first calculated the predicted velocity autocorrelation for the case that the dynamics of all three components are independent, $C_{\vec{v}, \text{indep}}(\tau) = C_s(\tau) C_{\psi}(\tau) C_{\Delta\psi}(\tau)$, where $C_s(\tau)$, $C_{\psi}(\tau)$, and $C_{\Delta\psi}(\tau)$ are the autocorrelation functions of the measured data for each of the components (see Supporting Information for derivation). We then compared the differences $C_{\vec{v},\text{obs}}-C_{\vec{v},\text{indep}}$ (blue curve in Figure \ref{Figure 4}C, bottom) and $C_{\vec{v},\text{obs}}-C_{\vec{v},\text{model}}$ (red curve in Figure \ref{Figure 4}C, bottom). Indeed, there were subtle differences both on shorter ($\sim$\SI{1}{\s}) and longer timescales ($\sim$\SI{10}{\s}). 
However, these errors for the simulated model were very similar to, or less than, those for the model-free prediction from the data under the assumption of independence (\textit{i.e.}, $C_{\vec{v},\text{obs}}-C_{\vec{v},\text{model}}\lesssim C_{\vec{v},\text{obs}}-C_{\vec{v},\text{indep}}$). These results demonstrate that modeling $s(t)$, $\psi(t)$, and $\Delta \psi(t)$ as independent stochastic processes provides a very good approximation to trajectory statistics across the ballistic-to-diffusive transition. The relatively subtle differences between the data and model arise primarily in instances where this assumption of independence between the three motility components breaks down. Consistent with these conclusions, inspection of cross-correlation functions computed from the data revealed that correlations between $s(t)$, $\psi(t)$, and $\Delta \psi(t)$ are largely absent, with only weak correlations between speed ($s$) and  reversals ($\Delta \psi$) in a subset of strains (Fig. S9).

\subsection*{Variation of exploratory behavior across Species}
The results presented in the previous sections demonstrate that a random-walk model with seven parameters describing independent speed, turning and reversal dynamics, provides a good approximation of the worms' motile behavior over the $\sim$\SI{100}{\second} timescale spanning the ballistic-to-diffusive transition. The model parameters thus define a seven-dimensional space of motility phenotypes in which behavioral variation across strains and species can be examined.  If components of behavior were physiologically regulated or evolutionarily selected for in a coordinated manner, we would expect to find correlated patterns in the variation of these traits.

We fit our model to the trajectory statistics of each individual worm and built a phenotype matrix of 106 worms x 7 behavioral parameters (summarized in Tables S2-S4). The correlation matrix for these 7 parameters demonstrates that the strongest correlation were the forward and reverse state lifetimes ($\tau_{fwd}$, $\tau_{rev}$), followed by those describing speed and forward state life times ($\mu_s$, $\tau_{fwd}$). More broadly, there were extensive correlations among the model parameters, not only within the parameters of each motility component (speed, orientation, reversals) but also between those of different components.

\begin{figure*}
\centerline{\includegraphics[scale=1]{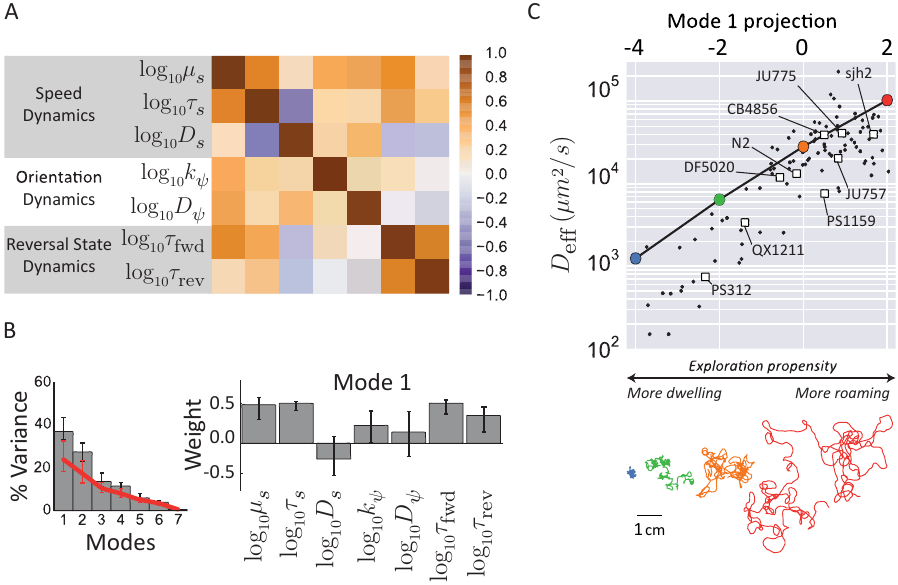}}
\caption{
{Motility parameters co-vary along an axis controlling exploratory behavior.}
(A) Correlation matrix of the behavioral parameters across the whole dataset.
(B, left) Fraction of variance captured by each mode and the amount expected for an uncorrelated dataset (red line).
(B, right) The loadings on the top eigenvector.
(C) The effective diffusivity (top) and a 30 minute trajectory (bottom, colors match points on graph) from simulations in which the loading on the top eigenvector was varied; the principal mode can be used as an effective phenotype from a more \emph{dwelling} to a more \emph{roaming} behavior. The projections and effective diffusivity of the measured trajectories are shown as black points, and the average of each strain is shown as a square.
}
\label{Figure 5}
\end{figure*}

We looked for dominant patterns in the correlations using principal component analysis \cite{P.Murphy2012} (Figure \ref{Figure 5}B), uncovering a single dominant mode of correlated variation (Figure \ref{Figure 5}B, left). 
This principal mode (mode 1), capturing nearly 40\% of the total variation, described significant correlations among all the parameters except for $D_s$ and $D_\psi$ (Figure \ref{Figure 5}B, right, Table S5). We did not attempt to interpret higher modes since, individually, they either did not significantly exceeded the captured variance under a randomization test (mode 3 and higher; see SI, and Figure \ref{Figure 5}B, left) or were found upon closer inspection to be dominated by parameter correlations arising from fitting uncertainties (mode 2).

We used numerical simulations to determine the effects on motile behavior of varying parameters along the principal mode. The measured trajectory phenotypes projected onto this mode in the range $\{-4, 2\}$ centered about the average phenotype located at the origin, and we performed simulations for parameter sets evenly sampled along this range. These largely reproduced the observed variation in the measured diffusivities $D_{\text{eff}}$ as a function of the projection along the first mode. The agreement was particularly good at  higher values ($>-1$) of the mode projection, but at lower values we noted a tendency for the $D_\text{eff}$ from simulations to exceed that of the data. The latter discrepancy can be explained by elements of behavior not captured by our model (see Discussion). Nevertheless, as illustrated by simulated trajectories (Figure \ref{Figure 5}C, bottom), trajectories became more expansive as the mode projection increased, as did $D_\text{eff}$ by nearly two orders of magnitude over the tested range. This suggested that the principal mode indicates exploratory propensity (Figure \ref{Figure 5}C), and we confirmed that it is indeed more strongly associated with changes in $D_\text{eff}$ than expected for randomly generated parameter sets (Figure S10). Interestingly, this mode of variation we found across individual phenotypes is reminiscent of ``roaming'' and ``dwelling'' behavioral variability that has been shown within individuals across time, in \textit{C. elegans}  \cite{Fujiwara2002,Flavell2013} as well as other organisms \cite{Osborne1997, Jordan2013}.

\subsection*{Specialized and Diversified Behavioral Strategies Across Strains}
The principal behavioral mode discussed in the preceding section was identified by analyzing variation across all individual worms measured in this study, coming from diverse strains and species that differ in their average behavior (see Tables S2 - S4). How does the variability among individuals of a given strain compare to differences between the average phenotypes of strains/species? On the one hand, each strain might be highly ``specialized'', with relatively small variation within strains as compared to that across strains. On the other hand, strains might implement ``diversified'' strategies in which genetically identical worms vary strongly in their behavior. To address these two possibilities, we analyzed the distribution of individual phenotypes within each strain, as well as that of the set of averaged species phenotypes.

\begin{figure*}
\centerline{\includegraphics[width=11.4cm]{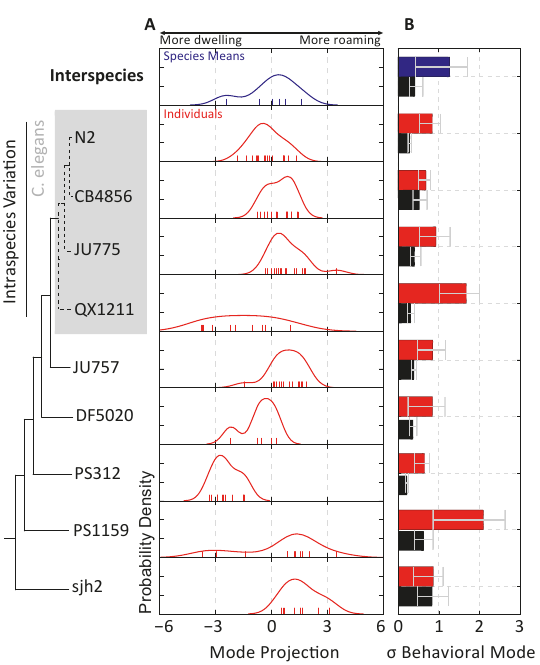}}
\caption{
{Variation of model parameters reveal specialized and diversified behavioral strategies across strains.}
(A) Distribution of the average phenotype for each species (\textit{interspecies} variation, blue) or individuals within a strain (red). Note that the variation of individual phenotypes in some strains (e.g. QX1211, PS1159) is comparable in magnitude to that of interspecies variation. Observations are indicated with colored ticks. (B) Comparison of the width of the phenotype distributions (colored bars), quantified as the bootstrapped standard deviation of the data points in (A), with the uncertainty in the determination of the phenotype (black bars), quantified as the standard deviation of individual phenotype determinations over bootstrapped \SI{100}{\second} time windows. Error bars correspond to 95\% confidence intervals across bootstrap samples.
}
\label{Figure 6}
\end{figure*}

For each measured individual, we computed the projection of its motility parameter set along the principal behavioral mode and estimated strain-specific distributions of this reduced phenotype (Figure \ref{Figure 6}, Table S6). In principle, any detail in the shape of these distributions could be relevant for evolutionary fitness, but here we focused our analysis on the mean and standard deviation, given the moderate sampling density ($\leq 20$ individuals per strain). Further, we computed the principal-mode projection of the average phenotype of each species to define an interspecies phenotype distribution (Figure \ref{Figure 6}).

Strains varied considerably in both the position and breadth of their phenotypic distributions along the principal behavioral mode. Remarkably, variation across individuals within each strain was comparable in magnitude to that for the set of average phenotypes across species (Figure \ref{Figure 6}). Some strains were specialized towards roaming or dwelling behavior, such as CB4856 and PS312, respectively, with a strong bias in their behavior and comparatively low individual variability. Others, such as QX1211 and PS1159, appeared more diversified with an intermediate average phenotype and higher individual variability. These considerable differences in phenotype distributions across strains reveal the evolutionary flexibility of population-level heterogeneity in nematodes, and suggest a possible bet-hedging mechanism for achieving optimal fitness in variable environments \cite{Slatkin1974,Philippi1989}.

In assessing such variability of phenotypes, it is essential to ask how uncertainty in the determined parameters (obtained from model fits) contribute to the observed variability in phenotypes. We therefore computed the contribution of uncertainties in the individual phenotype determination by bootstrap resampling of the \SI{100}{\second} windows of each individual's recorded trajectory (see SI). The uncertainties thus computed reflect contributions from both parameter uncertainties in curve fitting of data, as well as temporal variability in an individual's parameters over timescales longer than the window size (\SI{100}{\second}). With the exception of two strains (sjh2 and CB4856), this measure of uncertainty accounted for less than half of the individual variation within each strain (Figure \ref{Figure 6}B). These findings support the view that the phenotypic variation estimated in the current analysis largely represented stable differences in individual behavior.

\section*{Discussion}
We have presented a comparative quantitative analysis of motile behavior across a broad range of strains and species of the nematode phylum, ranging from the lab strain \textit{C. elegans} N2 to \textit{Plectus} sjh2 at the base of the chromadorean nematode lineage. Despite the vast evolutionary distances spanned by strains in this collection \cite{Kiontke2013}, we found that a behavioral model described by only seven parameters could account for much of the diversity of the worms' translational movement across the $\sim$\SI{100}{\second} timescale spanning the ballistic-to-diffusive transition. This simple model provides a basis for future studies aiming to capture more detailed aspects of nematode behavior, or to connect sensory modulation of behavior to the underlying physiology. More generally, our results demonstrate how quantitative comparisons of behavioral dynamics across species can provide insights regarding the design of behavioral strategies.

\subsection*{The Minimal Model: What Does It Capture, and What Does It Miss?}
We focused on a high-level output of behavior --- translational and orientational trajectory dynamics --- and sought to build the simplest possible quantitative model that could capture the observed behavioral statistics.  We found that a model with only three independent components --- (1) speed fluctuations that relax to a set point on a timescale of a few seconds, (2) orientation fluctuations with drift, and (3) stochastic switching between forward and reverse states of motion --- describes well, overall, the trajectory statistics of all tested nematode species across the ballistic-to-diffusive transition (Figure \ref{Figure 4}). 

Notably, we have not included explicit representations of some reorientation mechanisms that have been studied in the past, such as the deep turns (omega- and delta-turns) \cite{Croll1975,Broekmans2016}, or the combination of such turns  with reversals (pirouettes) \cite{Pierce-Shimomura1999}. In our data, we find that the timing of the initiation and termination of reversals, which would both count as runs in the pirouette description, follow exponential distributions with similar time constants as previously reported for the pirouette run distribution. While omega and delta turns must indeed be mechanistically distinct from gradual turns, we have chosen here not to explicitly model their occurrence since orientation changes in our trajectory data were adequately described by a continuous diffusion-drift process (Figures \ref{Figure 3}B, S6A). It is possible, however, that explicit representations of pirouettes and/or omega turns would be important in other experimental scenarios, e.g. those that include navigation in the presence of gradient stimuli.

In our model, "roaming" and "dwelling" were not assigned discrete behavioral states (as was done \textit{e.g.} in \cite{Fujiwara2002,Gallagher2013,Flavell2013}), but instead emerged as a continuous pattern of variation among motility parameters describing the worm's random walk. However, robust extraction of motility parameters required pre-filtering of trajectory data that likely biased them towards more "roaming" phenotypes (see SI), which we believe account for the noted tendency of model simulations to overestimate $D_\text{eff}$ that was more pronounced for trajectories at the "dwelling" end of the spectrum (Figure 5C). 

In its current form, our simple model does not account for possible correlations between the dynamics of the three motility components (speed, orientation, and reversals). Indeed, at least weak correlations do exist between the components (Figure S9). Comparisons of simulated versus measured trajectories demonstrated that the effects of such correlations on the motility statistics are small but detectable (Figure \ref{Figure 4}C). The differences were most significant for the velocity-autocorrelation dynamics on a $\sim$\SI{10}{\second} timescale, and were similar to those for model-free predictions obtained by combining component-wise correlation functions under the assumption of independence. Discrepancies on this intermediate timescale occurred most often in fast-moving strains that frequently approached the repellent boundary. Therefore, we suspect that the discrepancy arises from a stereotyped sequence, such as the escape response\cite{Culotti1978a}, that introduces temporal correlations between speed changes, turning, and reversals.

While here we have focused on the transition to diffusive motion, some recent experiments suggest that \textit{C. elegans} might engage in superdiffusive behavior on timescales longer than \SI{100}{\s} \cite{Peliti2013, Salvador2014}. Superdiffusive behavior could arise from nonstationarities in motile behavior, such as the roaming/dwelling transitions on timescales of several minutes \cite{Flavell2013}. Another mechanism for superdiffusion is \textit{directed} motility \cite{Peliti2013} in response to external stimuli such as chemical or thermal gradients.  In such environments, nematodes are known to use at least two distinct mechanisms for navigation \cite{Pierce-Shimomura1999,Iino2009a} and the model here could be extended by studying the dependence of motility parameters on environmental statistics.

Information about the body shape can be incorporated to build a more complete behavioural model that also includes dynamics hidden by centroid behaviour \cite{Stephens2008,Stephens2010}. Indeed, work by Brown et al. showed that a rich repertoire of dynamics can be identified as temporal ``motifs'' in the postural time series of \textit{C. elegans} and used to classify mutants with high discriminatory power\cite{Brown2013}. We have found that all of the species tested here can also be described with a common set of postural modes (not shown), suggesting future directions on the evolutionary space of postural dynamics.

\subsection*{The exploratory behavioral mode: Variability and its Physiological Basis}
While we found that a single behavioral model could be used to characterize nematode motility across the chromadorean lineage, the parameters of the model varied extensively from strain to strain. Quantitatively, about 37\% of the variation corresponded to a correlated change in the parameters underlying the timing of forward and reverse runs and the dynamics controlling speed and turning (Figure \ref{Figure 5}B). We find that this principal mode of variation is associated with strong changes in exploratory propensity, as characterized by $D_\text{eff}$ (Figure 5C). This pattern of parameter variation drove a change from low speed short runs to high speed long runs, resembling the canonical descriptions of roaming and dwelling in \textit{C. elegans} \cite{Flavell2013}.

Roaming and dwelling are thought to represent fundamental foraging strategies reflecting the trade-off between global exploration and local exploitation of environmental resources \cite{Davies2012}. Recent work has suggested that such archetypal strategies can be recovered by quantitatively analyzing the geometry of phenotypic distributions in parameter space \cite{Shoval2012,Gallagher2013}. The motility phenotypes we found in the present study were biased along one principal dimension, with the extremes corresponding to roaming and dwelling behaviors. This observation compels us to suggest that an exploration-exploitation trade-off is the primary driver of phenotypic diversification in the motility of chromadorean nematodes in the absence of stimuli. Interestingly, a recent study on the motility of a very different class of organisms (ciliates) yielded a similar conclusion\cite{Jordan2013}: across two species and different environments, the diversity of motility phenotypes was found to be distributed principally along an axis corresponding to roaming and dwelling phenotypes. The emergence of roaming/dwelling as the principal mode of variation in such disparate species underscores the idea that the exploration-exploitation trade-off is a fundamental constraint on biological motility strategies.

A surprising finding in our study was that, for a majority of strains, the extent of behavioral variability across individuals within a strain was comparable to that for variation of phenotypes across species (Figure \ref{Figure 6}). In slowly changing environments, the most evolutionarily successful species are those that consistently perform well in that environment. This can be achieved by evolving a specialized, high fitness phenotype that varies little among individuals (such as with PS312 and sjh2). However, increased phenotypic variability among  individuals can improve fitness in more variable environments if some individuals perform much better in each condition---a so-called ``bet-hedging'' strategy \cite{Slatkin1974,Philippi1989}. The large variability we observed among individual phenotypes within each strain might reflect such a bet-hedging strategy in nematode exploratory behavior. 

The observation that the variation among genetically identical individuals can be comparable to that between disparate species raises the intriguing possibility that there exist conserved molecular and/or physiological pathways driving diversification of spatial exploration strategies. Analogous variation in exploratory behavior was also detected in an analysis of nonstationarity in the behavior of wild-type and mutant \textit{C. elegans} under various nutritional conditions \cite{Gallagher2013}. Physiologically, protein kinase G (PKG) signaling and DAF-7 (TGF-$\beta$) signaling from the ASI neuron are thought to be major mechanisms controlling roaming and dwelling in \textit{C. elegans} \cite{Fujiwara2002,Gallagher2013}. PKG signaling is also involved in controlling foraging in \textit{Drosophila} and other insects as well as many aspects of mammalian behavior \cite{Reaume2009,Kaun2009}. Flavell et al. also elucidated a neuromodulatory pathway involving serotonin and the neuropeptide pigment dispersing factor (PDF) controlling the initiation and duration of roaming and dwelling states \cite{Flavell2013}.

Perturbations to the molecular parameters of such pathways underlying global behavioral changes might provide a mechanism for the observed correlated variations at the individual, intra-, and inter-species levels. The identification of such conserved pathways affecting many phenotypic parameters is of fundamental interest also from an evolutionary perspective, as they have been proposed to bias the outcome of random mutations towards favorable evolutionary outcomes \cite{Kirschner1998,Gerhart2007}. Our simple model provides a basis for future investigations to uncover conserved mechanisms that generate behavioral variability, by defining a succinct parameterization of behavior that can be combined with genetic and physiological methods.

\section*{Acknowledgments}
We thank Massimo Vergassola, Vasily Zaburdaev, Alon Zaslaver, and Jeroen van Zon for helpful suggestions and critical reading of the manuscript, Will Ryu, Aravi Samuel and Andre Brown for inspiration and encouragement, and members of the Shimizu lab for discussions. Casper Quist and Hans Helder of Wageningen University provided wild nematodes isolated from soil and useful information regarding the ecology of nematodes.

\clearpage

\bibliography{main}

\clearpage

\newcommand*\samethanks[1][\value{footnote}]{\footnotemark[#1]}

% Supplemental Figures and Tables
\beginsupplement
\section*{Supporting Information}
\section*{SI Materials and Methods}
\subsection*{Selection of Strains}
A phylogenetic tree with the strains used in this study is shown in Figure 1A. The nematode phylum is classically divided into three major \linebreak branches--chromadorea, enoplea, and dorylaimia--that are broken into a total of five major B-clades [S2]  and twelve minor H-clades [S3]. The chromadorean lineage is the largest, spanning B-clades III-V and H-clades 3-12 [S2,S3]. \textit{C. elegans} is located in clade V9 (the rhabditids), one of the most diverse clades [S4]. In addition to the lab strain N2, we selected three of the most genetically distinct wild isolates of \textit{C. elegans} (CB4856, JU775, and QX1211) to sample intraspecies variation [S5]. From H-clade 9 in order of increasing evolutionary distance, we selected \textit{Caenorhabditis briggsae} JU757, \textit{Rhabditis myriophila} DF5020, and \textit{Pristionchus pacificus} PS312. The next closest major group, B-clade IV, contains H-clades 10-12. H-clade 12 contains the plant parasitic tylenchs and was thus not included in this study. H-clades 10 and 11 contain many bacterial feeders, of which we selected \textit{Panagrolaimus sp.} PS1159. Finally, from the basal chromadorea, we obtained \textit{Plectus sp.} sjh2, a member of H-clade 6.

\textit{C. elegans} N2, CB4856 and JU775 were provided by the Caenorhabditis Genetics Center, which is funded by NIH Office of Research Infrastructure Programs (P40 OD010440). \textit{C. elegans} QX1211 was kindly provided by Erik Andersen (Northwestern Univ.). \textit{Plectus sp.} sjh2 was isolated from a soil sample using morphological criteria by Casper Quist and Hans Helder (Wageningen Univ.). SJH then isolated a single species by starting cultures with a single worm. The remaining strains were used in previous studies by Avery [S6].

\subsection*{Cultivation of Worms}
Worms were grown on NGM-SR plates (\SI{3}{\gram} NaCl, \SI{24}{\gram} agar, \SI{2.5}{\gram} peptone, \SI{1}{\milli\liter} \SI{5}{\milli\gram\per\milli\liter} cholesterol in EtOH in \SI{975}{\milli\liter} water, with \SI{1}{\milli\liter} \SI{1}{\Molar} CaCl$_2$, \SI{1}{\milli\liter} \SI{1}{\Molar} MgSO$_4$, \SI{25}{\milli\liter} \SI{1}{\Molar} K$_2$PO$_4$ pH 6, \SI{1}{\milli\liter} \SI{200}{\milli\gram\per\milli\liter} streptomycin in water, and \SI{0.23}{\gram} \SI{5}{\milli\liter} \SI{40}{\milli\gram\per\milli\liter} nystatin in DMSO, added after autoclaving) seeded with \textit{E. coli} HB101, as previously described [S7].  \textit{E. coli} HB101 was first cultured in M9 minimal media (\SI{3}{\gram} KH$_2$PO$_4$, \SI{6}{\gram} Na$_2$HPO$_4$, \SI{5}{\gram} NaCl, \SI{1}{\milli\liter} \SI{1}{\Molar} MgSO$_4$ in \SI{1}{\liter} water) supplemented with 10\% Luria broth and \SI{10}{\milli\gram\per\milli\liter} streptomycin [S8]. Plates were incubated with a light circle of HB101 culture for a day at \SI{37}{\degreeCelsius} and then stored at \SI{4}{\degreeCelsius}. For \textit{Plectus sp.} sjh2, low salt plates (2\% agar supplemented with \SI{5}{\milli\gram\per\liter} of cholesterol from a \SI{5}{\milli\gram\per\milli\liter} EtOH solution) were used as previously described [S9]. On NGM-SR plates, these worms became shriveled and died. As the plates did not have nutrients for the bacteria to grow, HB101 was grown to high density in Luria broth overnight at \SI{37}{\degreeCelsius}, washed 3X in water, resuspended at 10X concentration, and applied to the plates.

Nematodes were cultured by either transferring a few worms by worm pick or a chunk of agar to a new plate after the worms reached adulthood. The plates were then incubated at \SI{20}{\degreeCelsius}. The growth rate varied considerably among strains, with \textit{Plectus sp.} sjh2 taking nearly two weeks to reach adulthood. We avoided starving the worms at any point during their cultivation, especially in the period before behavioral experiments were performed, as this can induce transgenerational phenotypic changes [S10,S11], and we have observed transient effects on motility lasting at least a couple of generations (data not shown).

\subsection*{Imaging}
The imaging experiments were done on \SI{3.5}{\centi\meter} plates containing the same media used for cultivation. A 2$\times$2 \SI{10}{\milli\meter} repellant grid was made by etching the plate with a tool dipped in 1\% sodium dodecyl sufate, a detergent that \textit{C. elegans} and most other nematodes avoided. (Whereas many \textit{C. elegans} studies have used copper rings as a repellant boundary [S12], we found that it did not sufficiently repel other nematodes; data not shown). Four young adult, well-fed nematodes were transferred individually by worm pick to a \SI{10}{\micro\liter} drop of M9 (water for \textit{Plectus sp.} sjh2) to remove bacteria stuck to the worms. The worms were then transferred by pipette in a minimal amount of buffer to the imaging plate, and excess buffer was removed as much as possible. The plate was imaged 10-20 minutes after picking the worms, minimizing most transient behaviors. The plate was placed on a custom imaging rig in an inverted, uncovered configuration with illumination by a Schott MEBL-CR50 red LED plate. The behavior was recorded for 30 minutes using a Point Grey Grasshopper Express GX-FW-60S6M-C camera equipped with an Edmund Optics NT54-691 lens (set to a magnification of 0.5X) at a resolution of 2736x2192 (\SI{12.5}{\micro\meter\per px}) at \SI{11.5}{frames\per\second} using a custom National Instruments LabView acquisition program. The video was subsequently compressed using the open-source XVid MPEG-4 compression algorithm using maximal quality settings.

\subsection*{Tracking and Image Analysis}
The behavioral videos were analyzed using a custom automated analysis program in MathWorks Matlab. The average background was calculated from 50 frames evenly sampled across the entire video. The background was then subtracted from each frame and a global threshold was applied. The thresholded image was cleaned by applying a series of morphological operations: Incomplete thresholding of the worm was smoothed by applying morphological closing with a disk with a similar radius as the worm. Any remaining holes were filled in using a hole-filling algorithm. Small holes or ones with a low perimeter to area ratio were excluded as they sometimes fill in worms undergoing an omega turn, as described in [S13]. Finally, regions in which the worm was just barely touching itself were split by sequentially applying open, diagonal fill, and majority morphological operations. The worm was then identified as the largest connected component with an area within 2-fold of the expected value. The centroid was tracked across frames to obtain $\vec{x}(t)$. In addition, the image skeleton was calculated. Sample images from each of the processing steps are shown in Figure \ref{FigureS11}.

The head of the worm was automatically identified using two statistical properties of the worm's behavior, namely (i) on average, the head of the worm moves more than the tail, and (ii) on average, worms spend more time moving forward (in the direction of their head) than they do moving in reverse.  The procedure is based on skeletonization and centroid detection of the worm image, which can fail in situations where image contrast is low (e.g. due to non-uniform background), so trajectories were first divided into segments that contain no more than 3 frames missing the skeleton and centroid information, and the head orientation was assigned within each segment based on local behavioral statistics.  Finding statistical criteria that allow unambiguous assignment of head orientation across all strains studied here was challenging because of the diversity in their behavior, but the following procedure was found to work well empirically. The identity of the two ends of the skeleton across image frames were accounted for by a simple tracking algorithm based on minimizing the total distance between skeleton points. For segments longer than 150 frames (with no more than ten consecutive missing skeletons), we found that we could apply property (i) by computing the variance in body angles within 10\% of the body length from the ends, and assigning the head to the end with the greater summed variance. However, manual inspection revealed that this sporadicly resulted in misassignment of the head, identifiable as long reversals interrupted by short forward runs. Therefore, in addition, for segments longer than 200 frames (with no more than five consecutive missing centroids), we used property (ii), defining the head as the end of the skeleton that spent the majority of the trajectory at the leading edge of movement. Segments shorter than 150 frames were discarded from further analysis.

The velocity $\vec{v}(t)$ was calculated from the centroid position $\vec{x}(t)$ using the derivative of a cubic polynomial fit to a sliding \SI{1}{\second} window. The direct estimation of the velocity using a symmetrized derivative had a large $\delta$-correlated component that interfered with later analysis. The use of the cubic polynomial did not noticeably distort the correlation functions (Figure \ref{FigureS12}). When the worm's speed $s(t)=|\vec{v}(t)|$ is very low, its projections on the lab-frame x- and y-axes $v_x=\vec{v}(t)\cdot \hat{x}$ and $v_y=\vec{v}(t)\cdot \hat{y}$ become dominated by discretization (pixelation) noise, and the bearing $\phi(t)=\tan^{-1}(v_y/v_x)$ is poorly defined. This in turn leads to large fluctuations in $\Delta\psi(t)=\phi(t)-\psi(t)$, which can introduce a large number of false reversal events, noticeable as a steep decrease in the autocorrelation $C_{\Delta\psi}(\tau)=\langle\cos(\Delta\psi(t+\tau)-\Delta\psi(t))\rangle$ at small values of the delay $\tau$. We therefore exclude segments of the trajectories corresponding to run intervals shorter than six frames (less than half a second). When these artifacts are filtered out in this manner, the $\Delta\psi$ autocorrelation functions were well described by single exponentials (Figure \ref{FigureS7}C). We note that the exclusion of short runs effectively excludes segments of data in which the worm remains stopped (or at a very low speed) --- a feature that is more pronounced in some strains than others --- and this leads to a systematic bias for simulated model trajectories  to have a higher effective diffusivity $D_{\text{eff}}$ than the data for the corresponding strain (as can be seen in Figure 5C).

\subsection*{Calculation of Behavioral Statistics}
The worm's behavior fluctuated or sometimes drifted over long times (Figure \ref{FigureS4}), but the average statistics over \SI{100}{\second} windows were approximately stationary. In order to focus on dynamics within the \SI{100}{\second} timescale, the mean-squared displacement and all auto- and cross-correlation functions were calculated for \SI{100}{\second} windows and then averaged. This reduced the influence of longer timescale fluctuations in the speed and reversal rate. For all calculations, observations near the boundaries and pairs of points between which the worm approached the boundary were excluded. The uncertainty of each individual's phenotype projection on the principal behavioral mode was computed by projecting the motility parameters after bootstrapping over the 100s windows of each individual's trajectory. The standard deviation of the bootstrapped projections is used as uncertainty.

\subsection*{Calculation of Effective Diffusivity, $D_{\text{eff}}$}
To estimate the effective diffusivity $D_{\text{eff}}$, we fit the mean-squared displacement $\langle [\Delta x(\tau)]^2 \rangle$ over the diffusive regime. For this purpose, we defined the diffusive regime as the time-lag interval after which the normalized velocity autocorrelation $C_{\vec{v}}(\tau)/C_{\vec{v}}(0)$ decayed to below 0.1. We note that in some cases (especially for fast-moving strains such as CB4856, JU775 and sjh2) the fit to $\langle [\Delta x(\tau)]^2\rangle=4 D_{\text{eff}}\tau$ in this regime was poor due to boundary effects arising from the finite size of the behavioral arena. For these strains, $D_{\text{eff}}$ should be regarded as a lower bound for the true diffusivity.

\subsection*{Reversal Analysis}
The reversal state was assigned as described in the main text by analysis of $\Delta\psi(t)$. Assuming a random telegraph process that generates states $\Delta\psi$ = $0$ (forward) and $\Delta\psi$ = $\pi$  (reverse) with probabilities $1-f_{rev}$ and $f_{rev}$, respectively, the autocorrelation at long time lags is $C_{\Delta\psi}(\tau\to\infty) = (1 - 2 f_{rev})^2$.
For the proposed telegraph process, each state has an exponentially distributed lifetime ($\tau_\text{fwd}, \tau_\text{rev}$) and therefore $f_{rev} = \dfrac{\tau_\text{rev}}{\tau_\text{rev} + \tau_\text{fwd}}$. The expected correlation timescale for the mixture of the two states is $\tau_{RT}(\tau_\text{rev}, \tau_\text{fwd}) = \left( \tau_\text{fwd}^{-1} + \tau_\text{rev}^{-1} \right)^{-1} $. The $\Delta\psi$ autocorrelation function was therefore fit to
\begin{equation}
C_{\Delta\psi}(\tau) = \left[1-C_{\Delta \Psi \infty}(\tau_\text{rev}, \tau_\text{fwd})\right]\exp\left[-\dfrac{\tau}{\tau_{RT}(\tau_\text{rev}, \tau_\text{fwd})}\right]+C_{\Delta \Psi \infty}(\tau_\text{rev}, \tau_\text{fwd})
\end{equation}
where $C_{\Delta \Psi \infty}(\tau_\text{rev}, \tau_\text{fwd}) = (\dfrac{\tau_\text{fwd}-\tau_\text{rev}}{\tau_\text{fwd}+\tau_\text{rev}})^2$. The fraction of time spent reversing is: $f_\text{rev} = 0.5 - \sqrt{C_{\Delta \Psi \infty}(\tau_\text{rev},\tau_\text{fwd})/4}$, where $f_\text{rev} \in [0, 0.5]$. The transition time constants are then $\tau_\text{rev} = \dfrac{\tau_{RT}(\tau_\text{rev},\tau_\text{fwd})}{1-f_\text{rev}}$ and $\tau_\text{fwd} = \dfrac{\tau_{RT}(\tau_\text{rev},\tau_\text{fwd})}{f_\text{rev}}$.

To validate our approach, we compared the parameters obtained with our fitting procedure with those obtained from the distribution of time intervals between detected switching events (Figure S7). For both forward and reverse states, the distribution of time intervals between detected switching events (Figure S7B) were well-fit by a biexponential distribution $P(T_\text{run}>t)=C_{\Delta \Psi \infty} \\ \exp(-t/\tau_\text{short}) + (1-C_{\Delta \Psi \infty})\exp(-t/\tau_\text{long})$ with the time constants $\tau_\text{short}$ and $\tau_\text{long}$ typically separated by $>10$-fold, and the fraction of short intervals $C_{\Delta \Psi \infty}$ varying broadly over its full range, $0\le C_{\Delta \Psi \infty}\le1.0$ (Figure S7D,E). Values for $\tau_\text{short}$ were typically below \SI{1}{\second} (Figure S7D). While some fraction of these short intervals might represent true runs, they could also arise from spurious detection of switches in velocity bearing due to noise in estimating the centroid (see legend of Figure S7D) and in any event, contribute little to the overall dynamics of bearing decorrelation.

Values for $\tau_\text{fwd}$ and $\tau_\text{rev}$ obtained by fitting equation S1 to the measured autocorrelation functions correlated well with $\tau_\text{long}$ (Figure S7E), thus confirming that $\tau_\text{long}$ contributes to bearing randomization. We conclude that the forward/reverse switching dynamics are well described by equations (6) and (7), with parameters $\tau_\text{fwd}$, and $\tau_\text{rev}$.

\subsection*{Speed Analysis}
Transitions between forward and reverse runs tended to be excluded from the analysis because the speed crosses zero, rendering $\phi$ a noisy variable generating many short runs below our exclusion threshold of 6 frames (see above). The speed set point $\mu_s$ was fit by taking the mean. The remaining parameters of the speed dynamics (3) were fit by its analytical autocorrelation function: $C_s(\tau) = D_s \tau_s \exp\left(-\tau/\tau_s\right)$.

\subsection*{Orientation Analysis}
Changes in orientation during runs (\textit{i.e.} intervals between reversal events) were analyzed with respect to their mean-squared angular displacements (MSAD) over time, corresponding to a model for angular diffusion with drift.For an object lying on a two-dimensional plane, rotational diffusion about an axis normal to the plane leads to fluctuations in the orientation (an angle measured in the lab frame) $\psi(t)$ over time according to:
\begin{equation}\label{SIOrientationDiffusion}
d\psi(t) = \sqrt{2D_\psi} dW_t,
\end{equation}
where $D_\psi$ is the rotational diffusion coefficient, and $dW_t$ represents increments of a Wiener process. Bias in these fluctuations over time can be captured, to first order, by adding a linear drift term so that

\begin{equation}\label{SIOrientationDynamics}
d\psi(t) = k_{\psi} dt + \sqrt{2D_\psi} dW_t,
\end{equation}
with $k_\psi$ the drift coefficient.

If $k_\psi$ and $D_\psi$ are constant in time, the mean-squared angular displacement $\text{MSAD}(\tau) = \langle[\psi(t+\tau)-\psi(t)]^2\rangle$, is a quadratic function of the time delay $\tau$:
\begin{equation}\label{SIMSAD}
\langle[\psi(t+\tau)-\psi(t)]^2\rangle = \langle [k_{\psi} \tau + \sqrt{2D_\psi}(W_{t+\tau}-W_t)]^2\rangle =k_\psi^2 \tau^2 +2D_\psi\tau,
\end{equation}
where $\langle \cdot \rangle$ denotes averaging over all time pairs separated by $\tau$ and the last equality follows from the Wiener process properties $\langle W_{t+\tau}-W_t\rangle=0$ and $\langle [W_{t+\tau}-W_t]^2\rangle=\tau$.

More generally, if $k_\psi(t)$ and $D_\psi(t)$ are time-varying quantities, we can still approximate within a finite time window (centered about time $t_w$) the ``local'' values $k_{\psi,w}\approx k_\psi(t_w)$ and $D_{\psi,w}\approx D_\psi(t_w)$. In this study, we extract estimates of these (possibly time varying) parameters from fits to the averaged MSAD computed over time windows:

\begin{equation}\label{SIMSADW}
W^{-1}\sum_{w=1}^{w=W}{\langle[\psi(t+\tau)-\psi(t)]^2\rangle} =\langle k_{\psi}^2\rangle_w \tau^2 +2\langle D_\psi\rangle_w\tau,
\end{equation}
where $W$ is the number of windows and $\langle  x \rangle_w =W^{-1}\sum_{w=1}^{w=W}{x_{w}} $ represents averages over windows. By fitting this averaged MSAD by a quadratic function $a \tau +b \tau^2$, we thus obtain the estimates  $a/2=\langle D_\psi\rangle_w$ and $\sqrt{b}=\langle k_\psi^2 \rangle_w^{1/2}$. Note that $a/2$ obtained by this procedure yields an estimate of the mean value for $D_\psi$, but $\sqrt{b}$ corresponds to an estimate not of the mean value, but the root-mean-square (rms) value for $k_\psi$. Throughout the text,  we therefore explicitly refer to the latter estimate as $k_{\psi\mbox{rms}}$ (and refer to the former simply as $D_\psi$).

\subsection*{Simulations}
Reversals, orientation, and speed dynamics were all simulated independently using the model described. Forward and reverse run durations were chosen according to equations (5) and (6) by drawing exponential random numbers with mean value $\tau_\text{fwd}$ or $\tau_\text{rev}$. During reverse runs, $\Delta\psi$ was set to $\pi$. The orientation (4) and speed (3) dynamics were simulated using the Euler-Maruyama method [S14] %\cite{Kloeden1992}
with a time step that matched the frame rate. To prevent negative speeds, a reflective boundary condition was imposed by taking the absolute value of the speed at each simulation step. The velocity was then calculated from the decomposition in (1) and trapezoidally integrated to give the centroid position $\vec{x}(t)$.

\subsection*{Behavioral Mode Analysis}
The model parameters were fit to each trajectory to give a phenotypic matrix $\mathbf{T}$. The phenotypic matrix was centered by subtracting the mean phenotype, $\hat{\mathbf{T}} = \mathbf{T} - \langle\mathbf{T}\rangle_\text{indiv.}$. The correlation matrix was then calculated, $\mathbf{C_T} = \operatorname{corr}\hat{\mathbf{T}}$, and decomposed into eigenvalues $\lambda$ and eigenvectors (behavioral modes) $\mathbf{b}$, $\mathbf{C_T}\mathbf{b} = \lambda \mathbf{b}$. To reduce any bias coming from a single trajectory, this calculation was bootstrapped 1000 times. The significance of the $k$-th top mode is assessed by a comparison with the expected variance explained of the $k$-th top mode of randomly chosen directions in the behavioral space. We use the explained variance of the $k$-th mode of a newly created set of modes where the first $k-1$ modes are equal to the top behavioral modes and the remaining modes are pointing in randomly chosen orthogonal directions. This process is repeated 1000 times.

The projections of each trajectory on these behavioral modes were calculated by $\mathbf{P} = \hat{\mathbf{T}} \mathbf{b}$.
The uncertainty in the locus of each individual phenotype along the behavioral mode was computed by projecting the motility parameters after bootstrapping over the 100 second windows and taking the standard deviation. 

\subsection*{Statistics}
Unless otherwise indicated, errorbars and confidence intervals represent the 2.5\% and 97.5\% percentiles (spanning the 95\% confidence interval) estimated from 1000 bootstrap samples. All probability distributions were empirically estimated using kernel density methods in Python's Seaborn package with a bandwidth automatically selected using Scott's rule of thumb [S15]. Tabulated mean values of the effective diffusivity model and the motility model (Table \ref{Table S-MotilityParameters}-\ref{Table S-ReversalStateDynamicsParameters}) represent geometric rather than arithmetic means was used as the parameters varied log-normally.

\subsection*{Derivation of the Velocity Autocorrelation Function Under the Assumption of Independence}
The velocity autocorrelation function can be written in terms of the motility components,
\begin{equation}
\begin{split}
C_{\vec{v}}(\tau) & = \langle \vec{v}(0) \cdot \vec{v}(\tau) \rangle \\
 & = \langle s(0) \left[ \cos\left[\psi(0)+\Delta\psi(0)\right], \sin\left[\psi(0)+\Delta\psi(0)\right]\right]\times \\
 \times& s(\tau) \left[ \cos\left[\psi(\tau)+\Delta\psi(\tau)\right], \sin\left[\psi(\tau)+\Delta\psi(\tau)\right]\right]  \rangle 
\end{split}
\end{equation}

The expected value of the product of independent random variables is the product of the expected value of each variable, i.e. $\langle x y \rangle = \langle x \rangle \langle y \rangle$.
Therefore we can factor out $C_s = \langle s(0) s(\tau) \rangle$, leaving the vector product with $\psi$ and $\Delta\psi$.
The expanded vector product is:
\begin{equation}
\label{Cvindep_deriv1}
\begin{split}
C_{\vec{v}}(\tau) = C_s(\tau) \times \langle \cos \left[\psi(0) + \Delta\psi(0) \right] \cos \left[\psi(\tau) + \Delta\psi(\tau) \right]\\
+\sin \left[\psi(0) + \Delta\psi(0) \right] \sin \left[\psi(\tau) + \Delta\psi(\tau) \right] \rangle
\end{split}
\end{equation}
The trigonometric functions on $\psi(t) + \Delta\psi(t)$  can be rewritten as products of trigonometric functions of the terms:
\begin{eqnarray*}
\cos \left[ \psi(t) + \Delta\psi(t) \right] &=& \cos \psi(t) \cos \Delta\psi(t) - \sin \psi(t) \sin \Delta\psi(t) \\
\sin \left[ \psi(t) + \Delta\psi(t) \right] &=& \sin \psi(t) \cos \Delta\psi(t) + \cos \psi(t) \sin \Delta\psi(t)
\end{eqnarray*}
However, since $\Delta\psi(t) = \{ 0, \pi \}$, $\sin \Delta\psi(t) = 0$:
\begin{eqnarray*}
\cos \left[ \psi(t) + \Delta\psi(t) \right] &=& \cos \psi(t) \cos \Delta\psi(t)\\
\sin \left[ \psi(t) + \Delta\psi(t) \right] &=& \sin \psi(t) \cos \Delta\psi(t)
\end{eqnarray*}
Substituting into \eqref{Cvindep_deriv1},

\begin{equation}
\begin{split}
C_{\vec{v}}(\tau) = C_s(\tau) \times \langle  \cos  \psi(0) \cos \psi(\tau) \cos \Delta\psi(0) \cos \Delta\psi(\tau) + \\
\sin \psi(0)\sin\psi(\tau) \cos \Delta\psi(0) \cos \Delta\psi(\tau) \rangle
\end{split}
\end{equation}

We can now factor out $C_\psi(\tau) = \langle \cos \left[ \psi(\tau) - \psi(0)\right] \rangle =
\langle \cos \psi(0) \cos \psi(\tau) + \sin \psi(0) \sin \psi(\tau)\rangle$ to get:
\begin{equation*}
C_{\vec{v}}(\tau) = C_s(\tau) C_\psi(\tau) \langle \cos \Delta\psi(0) \cos \Delta\psi(\tau) \rangle
\end{equation*}
Finally, we substitute (again dropping $\sin \Delta\psi(t)$ terms):
\begin{equation*}
C_{\Delta\psi}(\tau) = \langle \cos\left[ \Delta\psi(0) - \Delta\psi(\tau) \right]\rangle =
\langle \cos\Delta\psi(0) \cos \Delta\psi(\tau)\rangle
\end{equation*} to get:
\begin{equation*}
C_{\vec{v}, \text{indep}}(\tau) = C_s(\tau) C_\psi(\tau) C_{\Delta\psi}(\tau)
\end{equation*}

\begin{figure*}
\centerline{\includegraphics[scale=1]{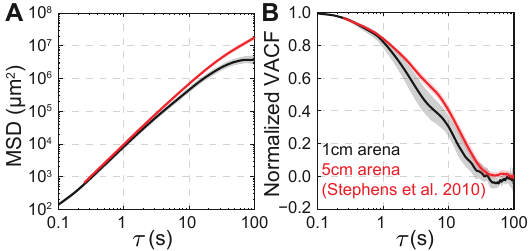}}
\caption{
Confinement by the boundary affects the mean-squared displacement (MSD) at long times, but does not impair resolution of the ballistic to diffusive transition.
We compare the statistical behavior of \textit{C. elegans} N2 in the experiments presented here within small (1-cm) arenas (black) and a previously reported dataset that used larger (5-cm) arenas [S1]  %\cite{Stephens2010}
(red). The MSD  (A), defined as $\langle [\Delta x(\tau)]^2 \rangle \equiv \langle |\vec{x}(t+\tau)-\vec{x}(t)|^2\rangle$, of our small-arena dataset is similar to that of the large-arena dataset at short times, but does show mild effects of confinement at long times ($\gtrsim \SI{100}{\second}$).
The ballistic to diffusive transition can be more closely studied by examining decay of the velocity autocorrelation function (VACF), defined as $C_v(\tau)\equiv \langle \vec{v}(t)\cdot\vec{v}(t+\tau)\rangle$ (B), which is related to MSD (\textit{i.e.} $[\Delta x(\tau)]^2$) by $(d/d\tau)\langle [\Delta x(\tau)]^2\rangle=2\int_{0}^{\tau}{d\tau' C_{\vec{v}}(\tau')}$ [S16]. The decay of the VACF to zero, which indicates orientation randomization and hence the transition from the ballistic to diffusive regime, is not significantly affected by the presence of the confining boundary.
}
\label{FigureS1}
\end{figure*}

\begin{figure*}
\centerline{\includegraphics[scale=1]{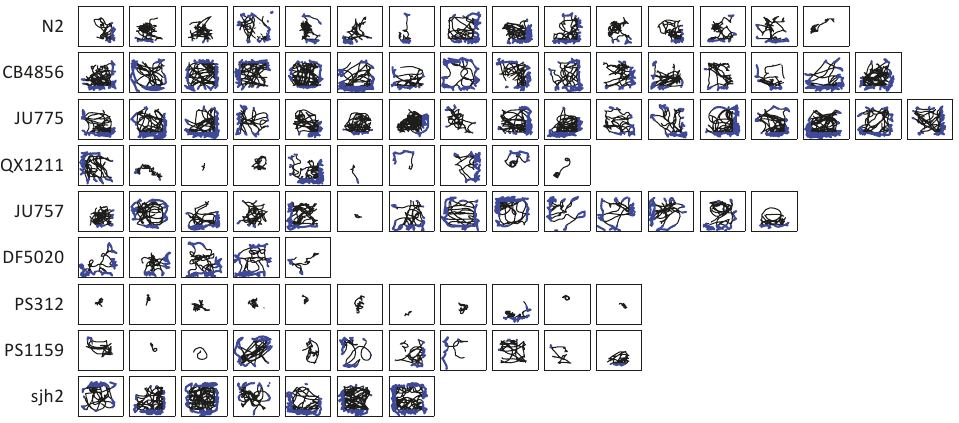}}
\caption{
{An overview of the dataset.}
Trajectories of all worm included in the study.
Each box represents a \SI{10}{\milli\meter} by \SI{10}{\milli\meter} chamber.
In blue, we highlight points excluded from the analysis because they were influenced by the boundary.
}
\label{FigureS2}
\end{figure*}

\begin{figure*}
\centerline{\includegraphics[scale=1]{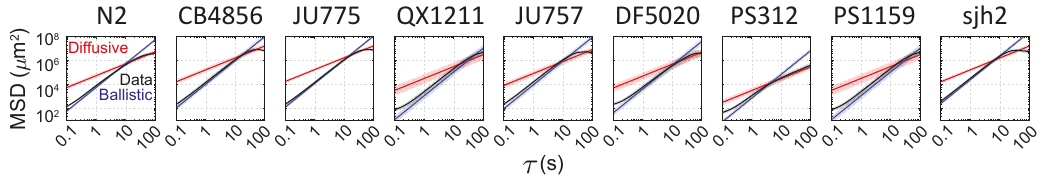}}
\caption{
{The ballistic to diffusive transition for all strains.}
We show the average mean-squared displacment (MSD), calculated across individual trajectories, for each strain (black). The expected ballistic (blue) and diffusive MSD curves (red), as in Figure 1B. 
}
\label{FigureS3}
\end{figure*}

\begin{figure*}
\centerline{\includegraphics[scale=1]{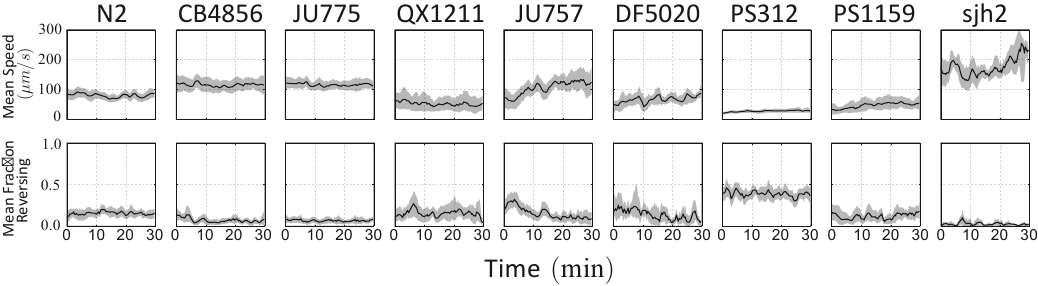}}
\caption{
{The worms' behavior was approximately stationary.}
For each strain, we show the average speed (top) and fraction of time spent reversing (bottom) calculated over \SI{100}{\second} sliding windows and averaged across individuals.
}
\label{FigureS4}
\end{figure*}

\begin{figure*}
\centerline{\includegraphics[scale=1]{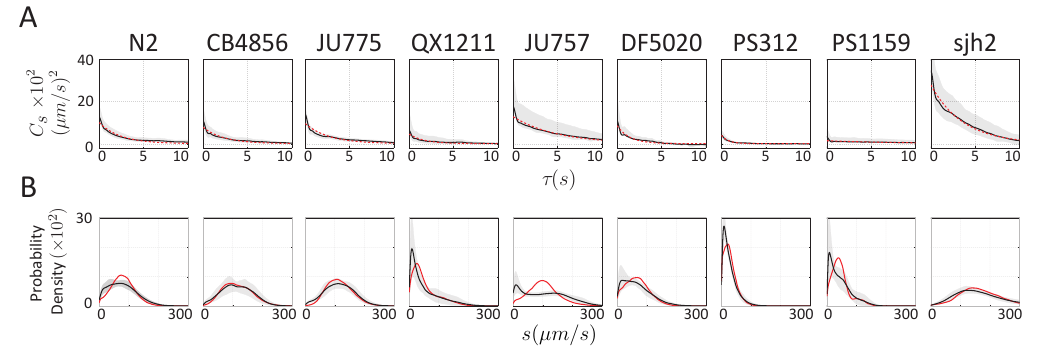}}
\caption{
{Characterization of speed statistics across strains.}
(A) The speed autocorrelation (black) of each strain decays exponentially (red).
(B) The speed distribution (black)  of each strain is closely reproduced by model simulations (red).
}
\label{FigureS5}
\end{figure*}

\begin{figure*}
\centerline{\includegraphics[scale=1]{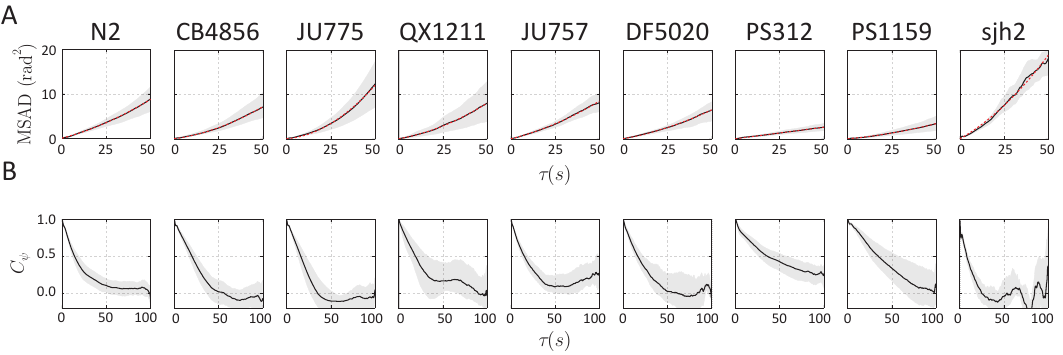}}
\caption{
{Characterization of orientation statistics across strains.}
(A) The mean squared angular displacement of body orientation (black) was fit to a quadratic function (red) in all strains.
(B) The orientation correlation (black) decays non-exponentially for many strains.
}
\label{FigureS6}
\end{figure*}

\begin{figure*}
\centerline{\includegraphics[width=11.4cm]{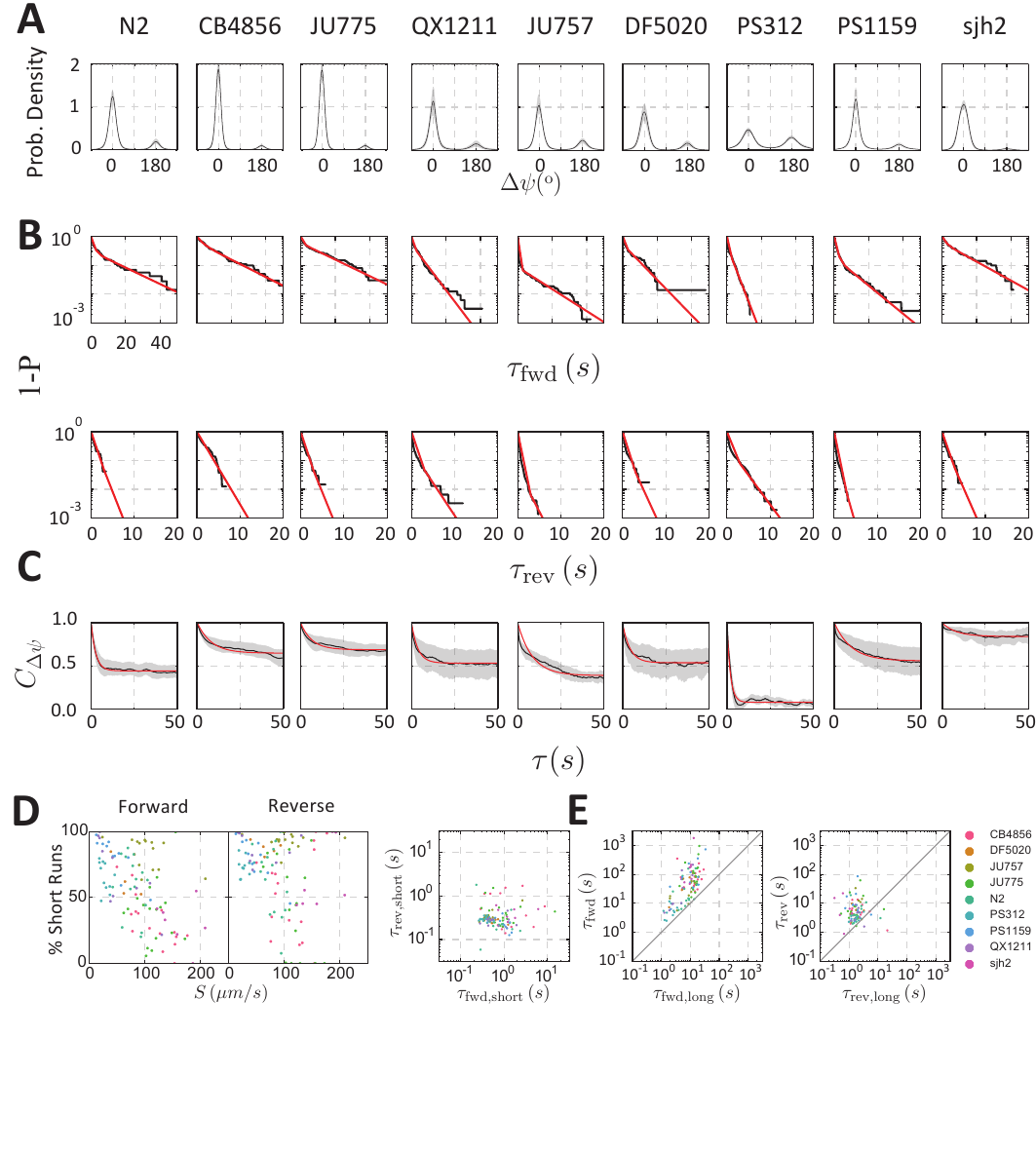}}
\caption{
{Characterization of reversal statistics across strains.}
(A) Distribution of $\Delta\psi$ for each strain shows two prominent peaks at $0\degree$ and $180\degree$.
(B) Cumulative distributions of the forward and reverse run durations ($T_\text{fwd}, T_\text{rev}$) for an individual worm from each strain (black), fit to a biexponential function (red). %As goodness-of-fit measures, $R^2$ and the Kullback-Leibler divergence $D_{KL}(\text{data}\|\text{fit})$ are computed for the interval $0<P<0.99$ and shown on each plot. 
(C) The autocorrelation function of $\Delta\psi$ for each strain (black) along with exponential fit (red).
(D, left) The fraction of short runs measured by the biexponential fits of the transition time distributions (as in B) was inversely correlated with the average speed of the worm. At low speed, the bearing (and therefore also $\Delta\psi$, which is used to identify runs) is expected to be dominated by noise (e.g. pixelation artifacts). (D, right) The fitted time constants for short forward and reverse intervals were uncorrelated (unlike those for long runs, see E and also Figure 5A), and typically below the timescale of smoothing filter for velocity data (\SI{1}{\second}), further motivating the exclusion of short intervals in modeling reversal dynamics.
(E) $\tau_\text{long}$, extracted from fits to the transition time distributions, were correlated with $\tau_\text{fwd}$ and $\tau_\text{rev}$, estimated from $C_{\Delta\psi}$ (panel C).
}
\label{FigureS7}
\end{figure*}

\begin{figure*}
\centerline{\includegraphics[scale=1]{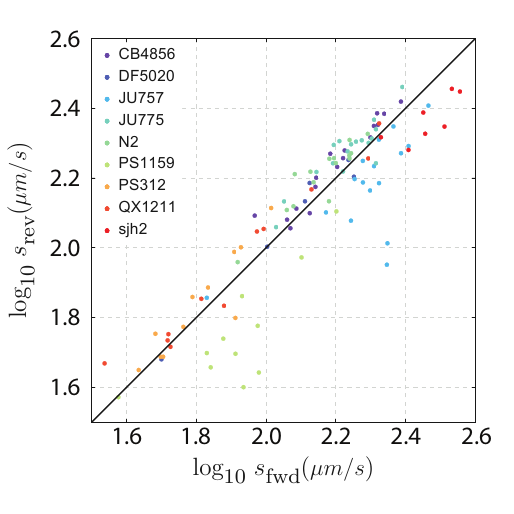}}
\caption{Worm run speeds are similar during forward and reverse runs for individual trajectories (dots, colored by strain). We define run speed as the top speed during runs (rather than the mean speed, to avoid biases due to run-length differences). The top speed is computed as the 95th percentile of the speed distribution (rather than the maximum, to avoid outlier effects). }
\label{FigureS8}
\end{figure*}

\begin{figure*}
\centerline{\includegraphics[scale=1]{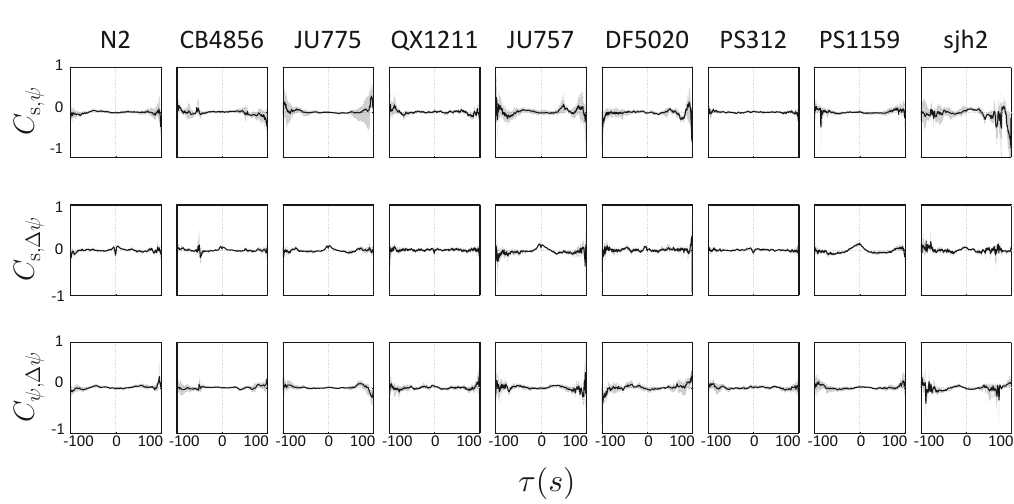}}
\caption{
{Cross-correlation analysis of motility dynamics.}
The cross-correlation between (top) speed and bearing changes, (middle) speed and velocity alignment, and (bottom) bearing changes and velocity alignment are shown for each strain. There is very little cross-correlation among the motility variables in any of the strains.  All cross-correlations were normalized to unit variance by dividing by the product of the standard deviation ($\sigma$) of the two components. 
}
\label{FigureS9}
\end{figure*}

\begin{figure*}
\centerline{\includegraphics[width=11.4cm]{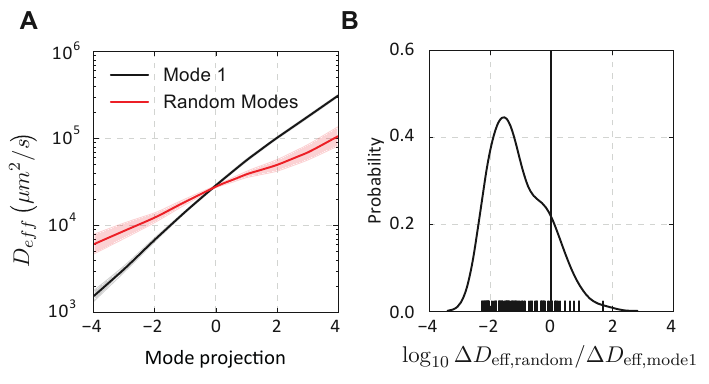}}
\caption{
{The top behavioral mode effectively captures changes in diffusivity compared to random projections.}
(A) The effect of variation on the top behavioral mode (black, as in Figure 5C) compared with a sampling of 100 random modes (red) on the diffusivity of simulated trajectories. For random modes, the sign of the mode was chosen such that the diffusivity increased with the projection along the mode. (B) For each random mode we compute the relative change in diffusivity between mode values $\Delta D_\text{eff}=D_\text{eff}(2)/D_\text{eff}(-2)$ and compare to the same relative diffusivity computed from the top behavioral mode. The kernel density distribution of the observed change is shown for the 100 samples (ticks). The black line indicates a ratio of 1 (no difference) and most random projections exhibit less range in  $\Delta D_\text{eff}$.}
\label{FigureS10}
\end{figure*}

\begin{figure*}
\centerline{\includegraphics[width=8.7cm]{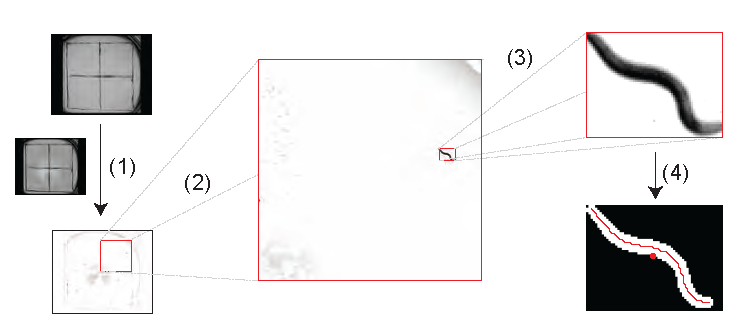}}
\caption{
{An Overview of the image processing steps.}
The video frames were processed by (1) subtracting the average of 50 frames evenly sampled from the entire movie and (2) cropping to each of the SDS-enclosed regions.
(3) The largest worm-sized object was identified following several image morphology operations, and (4) the centroid and image skeleton were measured.
}
\label{FigureS11}
\end{figure*}

\begin{figure*}
\centerline{\includegraphics[width=11.4cm]{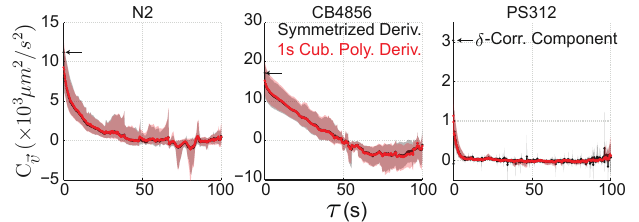}}
\caption{
{Comparison of velocity calculation methods.}
Velocity autocorrelation functions for the three example strains with and without filtering of the data and without averaging over \SI{100}{\second} windows.
The unfiltered velocity (black), estimated using a symmetrized derivative, contained a $\delta$-correlated short-timescale component in all strains that was particularly prominent in slow-moving strains such as PS312.
The velocity calculated using a \SI{1}{\second} cubic polynomial filter (red) does not contain this $\delta$-correlated component.
}
\label{FigureS12}
\end{figure*}

\clearpage

\begin{table}
\centering
\begin{tabular}{l|rrr}%|rrr}
 &
\multicolumn{3}{c}{$D_\text{eff}\times 10^2$} \\%&

 &
\multicolumn{3}{c}{$(\mu m^2/s) $} \\%&

Strain &
Mean & 2.5\% & 97.5\% \\%&
\hline
 N2        &                140 &  105 &   185 \\
 CB4856    &          429	&	307	& 620 \\
 JU775     &            448 & 360 & 558 \\
 QX1211    &          36 & 12 & 98 \\
 JU757     &            210 & 123 & 327 \\
 DF5020    &        128 & 55 & 255 \\
 PS312     &          8 & 5 & 13\\
 PS1159    &        81 & 32 & 183 \\
 sjh2      &           425 & 314 & 553 \\

\end{tabular}
\caption{The geometric mean of the effective diffusivity for each strain. For each trajectory, an effective diffusivity ($D_\text{eff}$) was extracted by analysis of mean-squared displacements and the velocity autocorrelation function.}
\label{Table S-MotilityParameters}
\end{table}

\begin{table}
\centering
\begin{tabular}{l|rrr|rrr|rrr}
 & \multicolumn{3}{c|}{$\mu_s$} &  \multicolumn{3}{c|}{$\tau_s$} &  \multicolumn{3}{c}{$D_s\times 10^2$}  \\
 & \multicolumn{3}{c|}{$\left( \mu m /s \right) $} &  \multicolumn{3}{c|}{$\left( s \right)$} &  \multicolumn{3}{c}{$((\mu m/s)^2 /s)$}  \\
 Strains   &   Mean &   2.5\% &   97.5\% &   Mean &   2.5\% &   97.5\% &   Mean &   2.5\% &   97.5\% \\
\hline
 N2        &          77 &     68 &      85 &            1.9 &      1.5 &       2.4 &        5.8 &    4.3 &     7.9 \\
 CB4856    &         108 &     91 &     128 &            1.8 &      1.5 &       2.3 &        4.0 &    3.1 &     5.2 \\
 JU775     &         112 &     98 &     127 &            2.1 &      1.6 &       2.7 &        4.3 &    3.3 &     5.5 \\
 QX1211    &          40 &     26 &      66 &            0.7 &      0.4 &       1.1 &        7.5 &    4.6 &    11.6 \\
 JU757     &          97 &     72 &     120 &            4.2 &      3.3 &       5.2 &        3.8 &    2.8 &     5.0 \\
 DF5020    &          65 &     50 &      83 &            1.1 &      0.8 &       1.3 &       11.7 &    9.1 &    14.4 \\
 PS312     &          27 &     23 &      32 &            0.7 &      0.6 &       0.8 &        5.3 &    3.7 &     7.4 \\
 PS1159    &          38 &     26 &      53 &            3.1 &      1.4 &       6.2 &         0.6 &     0.3 &     1.1 \\
 sjh2      &         159 &    138 &     184 &            3.3 &      2.5 &       4.7 &        8.8 &    5.8 &    13.2 \\
\end{tabular}
\caption{The model parameters related to the speed dynamics are listed for each strain. For each worm in a strain, time-averaged parameters were calculated.}
\label{Table S-SpeedDynamicsParameters}
\end{table}

\begin{table}
\centering
\begin{tabular}{l|rrr|rrr}

 & \multicolumn{3}{c|}{$k_{\psi \mbox{rms}}$} &  \multicolumn{3}{c}{$D\psi$} \\
 & \multicolumn{3}{c|}{$\left( rad/s \right) $} &  \multicolumn{3}{c}{$\left( rad^2/s \right)$}  \\
 Strains   &   Mean  &   2.5\% &   97.5\% &   Mean &   2.5\% &   97.5\% \\
\hline
 N2        &        0.036 &  0.026 &   0.048 &        0.034 &  0.017 &   0.054 \\
 CB4856    &        0.029 &  0.018 &   0.041 &        0.024 &  0.018 &   0.033 \\
 JU775     &        0.038 &  0.026 &   0.053 &        0.021 &  0.016 &   0.030 \\
 QX1211    &        0.040 &  0.028 &   0.056 &        0.017 &  0.009 &   0.036 \\
 JU757     &        0.039 &  0.030 &   0.052 &        0.036 &  0.021 &   0.054 \\
 DF5020    &        0.037 &  0.032 &   0.042 &        0.033 &  0.026 &   0.041 \\
 PS312     &        0.017 &  0.011 &   0.029 &        0.021 &  0.014 &   0.028 \\
 PS1159    &        0.023 &  0.015 &   0.031 &        0.009 &  0.005 &   0.017 \\
 sjh2      &        0.066 &  0.057 &   0.077 &        0.090 &  0.065 &   0.127 \\

\end{tabular}
\caption{The model parameters related to the orientation dynamics are listed for each strain. For each worm in a strain, time-averaged parameters were calculated.}
\label{Table S-OrientationDynamicsParameters}
\end{table}

\begin{table}
\centering
\begin{tabular}{l|rrr|rrr}
 & \multicolumn{3}{c|}{$\tau_{\mbox{fwd}}$} &  \multicolumn{3}{c}{$\tau_{\mbox{rev}}$} \\
 & \multicolumn{3}{c|}{$\left( s \right) $} &  \multicolumn{3}{c}{$\left( s \right)$}  \\
 Strains   &   Mean  &   2.5\% &   97.5\% &   Mean &   2.5\% &   97.5\% \\
\hline
 N2        &           23.8 &   13.9 &    41.1 &            4.1 &    3.0 &     5.7 \\
 CB4856    &           78.6 &   56.5 &   109.5 &            4.3 &    2.8 &     6.3 \\
 JU775     &           85.3 &   51.5 &   144.0 &            8.0 &    4.2 &    16.3 \\
 QX1211    &           26.5 &   12.5 &    63.0 &            3.7 &    2.7 &     5.2 \\
 JU757     &           32.3 &   20.5 &    50.5 &            6.4 &    4.1 &    10.3 \\
 DF5020    &           32.7 &   15.2 &    63.2 &            4.8 &    3.2 &     7.2 \\
 PS312     &            5.6 &    4.2 &     7.4 &            3.3 &    2.9 &     3.9 \\
 PS1159    &           80.8 &   38.0 &   174.6 &            8.8 &    5.2 &    16.2 \\
 sjh2      &          155.5 &   75.2 &   419.9 &            4.9 &    2.3 &    10.8 \\
\end{tabular}

\caption{The model parameters related to the reversal state dynamics are listed for each strain. For each worm in a strain, time-averaged parameters were calculated.}
\label{Table S-ReversalStateDynamicsParameters}
\end{table}

\begin{table}
\centering
\begin{tabular}{l|rrr}
& \multicolumn{3}{c}{Loading} \\
 Parameter                 &   Mean &   2.5\% &   97.5\% \\
\hline
 $\log_{10} \mu_s $   &        0.50 &   0.25 &    0.58 \\
 $\log_{10} \tau_s $            &           0.51 &   0.40 &    0.54 \\
 $\log_{10} D_s $ &          -0.19 &  -0.43 &    0.08 \\
 $\log_{10} k_{\psi}$      &           0.24 &  -0.04 &    0.43 \\
 $\log_{10} D_{\psi} $    &           0.15 &  -0.22 &    0.40 \\
 $\log_{10}\tau_{fwd}$        &           0.50 &   0.36 &    0.55 \\
 $\log_{10}\tau_{rev}$        &            0.35 &   0.15 &    0.48 \\
\end{tabular}

\caption{The loadings of each parameter on the top behavioral mode are listed.}
\label{Table S-BehavioralModeLoadings}
\end{table}

\begin{table}
\centering

\begin{tabular}{l|rrr}
& \multicolumn{3}{c}{Projection}\\
 Strain   &   Mean &   2.5\% &   97.5\% \\
\hline
 N2       &             -0.26 &  -0.48 &    0.49 \\
 CB4856   &              0.42 &  -0.35 &    0.76 \\
 JU775    &              0.86 &  -0.06 &    1.04 \\
 QX1211   &             -1.45 &  -1.82 &    0.32 \\
 JU757    &              0.78 &  -0.07 &    1.23 \\
 DF5020   &             -0.65 &  -0.70 &    1.19 \\
 PS312    &             -2.42 &  -2.66 &   -0.13 \\
 PS1159   &              0.45 &  -2.64 &    1.01 \\
 sjh2     &              1.61 &   0.80 &    2.41 \\

\end{tabular}

\caption{The phenotypic projection along the first behavioral mode is listed for each strain. For each worm in a strain, a time-averaged projection was calculated.}
\label{Table S-StrainBehavioralModeProjections}
\end{table}

\clearpage

\renewcommand\refname{SI references}

\end{document}